\documentclass[onecolumn,showpacs,preprintnumbers,superscriptaddress,pre]{revtex4}
\usepackage{subfigure}
\usepackage{graphicx,dcolumn,bm,color,latexsym,amssymb}
\usepackage[latin1]{inputenc}

\definecolor{Blue}{rgb}{0.0,0.0,1}
\definecolor{Red}{rgb}{1,0.0,0.0}
\definecolor{Green}{rgb}{0,0.5,0.0}

\begin{document}

\title{Finite size analysis of a two-dimensional Ising model within a
nonextensive approach}
\author{N. Crokidakis}
\email{nuno@if.uff.br}
\affiliation{Instituto de Física - Universidade Federal Fluminense, Av. Litor\^anea s/n,
24210-340 Niter\'oi - RJ, Brazil.}
\author{D.O. Soares-Pinto}
\email{dosp@cbpf.br}
\affiliation{Centro Brasileiro de Pesquisas Físicas, Rua Dr. Xavier Sigaud 150, Urca,
22290-180 Rio de Janeiro - RJ, Brazil.}
\author{M.S. Reis}
\affiliation{CICECO, Universidade de Aveiro, 3810-193 Aveiro, Portugal.}
\author{A.M. Souza}
\affiliation{Institute for Quantum Computing and Department of Physics and Astronomy,
University of Waterloo, Waterloo, Ontario, N2L 3G1, Canada.}
\author{R.S. Sarthour}
\author{I.S. Oliveira}
\affiliation{Centro Brasileiro de Pesquisas Físicas, Rua Dr. Xavier Sigaud 150, Urca,
22290-180 Rio de Janeiro - RJ, Brazil.}
\date{\today}
\date{\today}

\begin{abstract}
\noindent In this work we present a thorough analysis of the phase
transitions that occur in a ferromagnetic 2D Ising model, with only
nearest-neighbors interactions, in the framework of the Tsallis nonextensive
statistics. We performed Monte Carlo simulations on square lattices with
linear sizes L ranging from 32 up to 512. The statistical weight of the
Metropolis algorithm was changed according to the nonextensive statistics. 
Discontinuities in the m(T) curve are observed for $q\leq 0.5$. 
However, we have verified only one peak on the
energy histograms at the critical temperatures, indicating the occurrence of
continuous phase transitions. For the $0.5<q\leq 1.0$ regime, we have found
continuous phase transitions between the ordered and the disordered phases,
and determined the critical exponents via finite-size scaling. We verified
that the critical exponents $\alpha $, $\beta $ and $\gamma $ depend on the
entropic index $q$ in the range $0.5<q\leq 1.0$ in the form $\alpha
(q)=(10\,q^{2}-33\,q+23)/20$, $\beta (q)=(2\,q-1)/8$ and $\gamma
(q)=(q^{2}-q+7)/4$. On the other hand, the critical exponent $\nu $ does not
depend on $q$. This suggests a violation of the scaling relations $2\,\beta
+\gamma =d\,\nu $ and $\alpha +2\,\beta +\gamma =2$ and a nonuniversality of
the critical exponents along the ferro-paramagnetic frontier.
\end{abstract}

\maketitle


\section{Introduction}

Inspired in the geometrical theory of multifractals, Tsallis has suggested a
generalization of the Boltzmann-Gibbs entropy ($S_{BG}$), which is known as
the nonadditive entropy \cite{1988_JSP_52_479, livro_tsallis}. The entropy
form is postulated to be 
\begin{equation}
S_{q}=k\frac{1-\sum_{i}p_{i}^{q}}{q-1}~,  \label{eq.01}
\end{equation}%
\noindent where $\sum_{i}p_{i}=1$ and $k$ is a constant. The idea behind
this generalization is that $S_{q}$ is the measure of the information of
biased systems. Thus, being $p_{i}$ the probability of finding a given
system on the state $i$, the factor $q$ introduces a bias into the
probability set, i.e., if $0<p_{i}<1$ then $p_{i}^{q}>p_{i}$ for $q<1$ and $%
p_{i}^{q}<p_{i}$ for $q>1$. In other words, $q<1$ privileges the less
probable events in opposition to the more probable ones, and vice-versa.
This entropy is invariant under permutations, becomes zero for the maximum
knowledge about the system, and for $q=1$, i.e. for unbiased systems, it
would recover the BG entropy. The bias factor $q$ is called the entropic
index and $q\in \Re $. Recently, it has been proposed that $q$ is connected
to the dynamics of the system \cite%
{2002_PRE_66_045104,2004_PhysA_340_1,2005_EPL_71_339,2006_PRB_73_092401,2006_EPJB_50_99,2008_BJP_38_203}%
. Besides representing a generalization, the nonextensive entropy $S_{q}$,
as much as $S_{BG}$, is positive, concave and Lesche-stable ($\forall q>0$).
It has also been shown that for systems with certain types of correlations
that induces scale invariance in the phase space, the entropy $S_{q}$
becomes additive \cite%
{website_TEMUCO,2005_PNAS_102_15377,2008_JSMTE_09_006,2008_PRE_78_021102}.
The optimization of the entropy in Eq.(\ref{eq.01}) leads to the equilibrium
distribution and a generalization of the Boltzmann-Gibbs statistics, that is
called nonextensive statistics. This generalization has been successfully
applied in many areas of physics, biology and computation in the past few
years \cite{2006_dosp_bio,2007_silvio_epjb,2006_pre_boghosian,
2007_jstat_dosp}.

On the other hand, magnetic models are one of the most studied systems in
condensed matter, and the Ising model is a prototype that has been
extensively investigated for the last 30 years. More recently, some works
have investigated the magnetic properties of some manganese oxides, called
manganites, and connections with the nonextensive statistics have been
proposed \cite%
{2006_PRB_73_092401,2002_PRB_66_134417,2002_EPL_58_42,2003_PRB_68_014404}.
In a recent work, Reis \textit{et al.} \cite{2003_PRB_68_014404} studied the
phase transitions that occur in a classical spin system within the
mean-field approximation, in the framework of Tsallis nonextensive
statistics, and some interesting properties were found: the system presents
first-order phase transitions for $q<0.5$, but only continuous transitions
were found for $0.5<q<1.0$. The results present qualitative agreement with
experimental data in the $\mathrm{La_{0.60}Y_{0.07}Ca_{0.33}MnO_{3}}$
manganite \cite{2002_jmmm_amaral}.

From this, two natural questions appear: What properties of
infinite-range-interaction models can appear in short-range-interaction
ones? Can the mean-field predictions be verified on low dimensional systems?
Although a 2D Ising model, defined in the limit of nearest-neighbors
interactions, is usually treated according the BG statistics, a nonextensive
approach can be seen as a toy model for the elucidation of the question:
will the mean-field behavior be also present in a short-range-interaction
model? Furthermore, using this kind of study one can also verify the
accuracy of the nonextensive statistics applied to magnetic systems.

In a attempt to clarify some of these questions, we report in this paper
results on the study of the phase transitions that occur on a 2D Ising model
within a nonextensive approach. These results were obtained through Monte
Carlo (MC) simulations upon replacing the statistical weight of the
Metropolis algorithm by the nonextensive one. We performed a finite size
scaling in order to estimate the critical exponents for different values of
the entropic index $q \in [0,1]$. As discussed in Ref.\cite{2008_EPJB_62_337}%
, the critical temperatures $T_{c}$ depend on $q$ ($\forall\,q \in [0,1]$),
but we have found in the present work that the critical exponents $\alpha$, $%
\beta$ and $\gamma$ depend on the entropic index $q$, in the range $%
0.5<q\leq 1.0$.

\section{Nonextensive statistics and Monte Carlo simulation}

In nonextensive statistics theory (see, e.g., Refs.\cite%
{livro_tsallis,1998_PhysA_102_15377,livro_TsallisCap1} for details), the
energy constraint is given by 
\begin{equation}
\langle \mathcal{H}\rangle _{q}\equiv \sum_{i=1}^{\Omega }P_{i}\,\varepsilon
_{i}=U_{q}~,  \label{eq.02}
\end{equation}%
\noindent in which $\mathcal{H}$ is the hamiltonian of the system under
consideration, $\varepsilon _{i}$ represent the $\Omega $
possible energy states, and we have introduced the concept of escort
distribution \cite{livro_beck} 
\begin{equation}
P_{i}\equiv \frac{p_{i}^{q}}{\sum_{j=1}^{\Omega }p_{j}^{q}}=\frac{\left[
e_{q}^{-\beta _{q}^{\prime }\,\varepsilon _{i}}\right] ^{q}}{%
\sum_{j=1}^{\Omega }\left[ e_{q}^{-\beta _{q}^{\prime }\,\varepsilon _{j}}%
\right] ^{q}},  \label{eq.03}
\end{equation}%
\noindent where 
\begin{equation}
\beta _{q}^{\prime }=\frac{\beta }{\sum_{j=1}^{\Omega
}p_{j}^{q}+(1-q)\,\beta \,U_{q}},  \label{eq.04}
\end{equation}%
\noindent being $\beta $ the Lagrange parameter associated with the
constraint in Eq.(\ref{eq.02}), $e_{q}^{x}\equiv \left[ 1+(1-q)\,x\right]
_{+}^{1/(1-q)}$ the $q-$exponential and $[y]_{+}\equiv y\,\theta (y)$, where 
$\theta (y)$ denotes the Heaviside step function. This implies a cutoff for $%
\varepsilon _{i}$ given by 
\begin{equation}
\beta _{q}^{\prime }\,\varepsilon _{i}<\frac{1}{1-q}.  \label{eq.06}
\end{equation}%
\noindent The definition of the physical temperature in the nonextensive
statistics is still an open issue \cite%
{1999_PhysA_268_553,2000_PhysA_283_59,2001_PhysA_295_246,
2001_PhysA_295_416,2001_PLA_281_126,2002_PhysA_305_52,2003_PhysA_317_209,2004_EPL_65_606,2006_PhysA_368_430}%
. From a pragmatic point of view, since $(\beta _{q}^{\prime })^{-1}$ has
the dimension of energy, $(\beta _{q}^{\prime }k)^{-1}$ is a temperature
scale which can be used to interpret experimental results. The validity of
this choice was first shown experimentally \cite{2002_EPL_58_42}, and later
theoretically \cite%
{2006_PRB_73_092401,2006_EPJB_50_99,2002_PRB_66_134417,2003_PRB_68_014404}
for manganites.

The MC technique has been successfully used to study the physical properties
of Ising models \cite{1995_PhysA_222_205,1997_PhysA_244_344}. Thus, in order
to generalize the study of the properties of this model by a nonextensive
approach, we modified the Metropolis method for the nonextensive statistics.
To proceed the single spin flip MC calculations \cite{livro_compphys} and to
obtain the physical quantities of interest of the system (magnetization,
susceptibility, specific heat, and other quantities), we have changed the
usual statistical weight to \cite{2008_EPJB_62_337} 
\begin{equation}
w_{q}=\frac{P_{i,after}}{P_{i,before}}=\left[ \frac{e_{q}^{-\varepsilon
_{i}^{after}/k\,T}}{e_{q}^{-\varepsilon _{i}^{before}/k\,T}}\right] ^{q}~.
\label{eq.07}
\end{equation}%
\noindent The above equation is the ratio between the escort probabilities,
Eq.(\ref{eq.03}), before and after the spin flip, and $\varepsilon _{i}$ are
the energy states related to the Hamiltonian: 
\begin{equation}
\mathcal{H}=-J\sum_{\langle ij\rangle }s_{i}s_{j},  \label{eq.08}
\end{equation}%
\noindent where $\langle ij\rangle $ denotes the sum over nearest-neighbors
on a square lattice of size $N=L^{2}$, $s_{i}=\pm 1$ and $J>0$ (which
implies a ferromagnetic interaction).

Since Eq.~(\ref{eq.07}) is a ratio, the calculated weight can also be
written as the ratio between the $q-$exponentials with a bias $q$ \cite%
{2008_EPJB_62_337}. It is important to emphasize that $w_{q}$ is the
quantity that will be compared to a random number in the Metropolis
algorithm and also to note that the cutoff procedure, Eq.(\ref{eq.06}), must
be taken into account, i.e., it must be included into $w_{q}$ to avoid
complex probabilities.

Taking into account Eqs.~(\ref{eq.06}), (\ref{eq.07}), and (\ref{eq.08}), we
performed MC simulations with the entropic index $q\in \lbrack 0,1]$ and
linear lattice sizes of $L=32,64,128,256$ and $512$, with periodic boundary
conditions and a random initial configuration of the spins. The following
results were obtained after $10^{7}$ MC steps per spin.
 
\begin{figure}[t]
\begin{center}
\includegraphics[width=0.32\textwidth,angle=0]{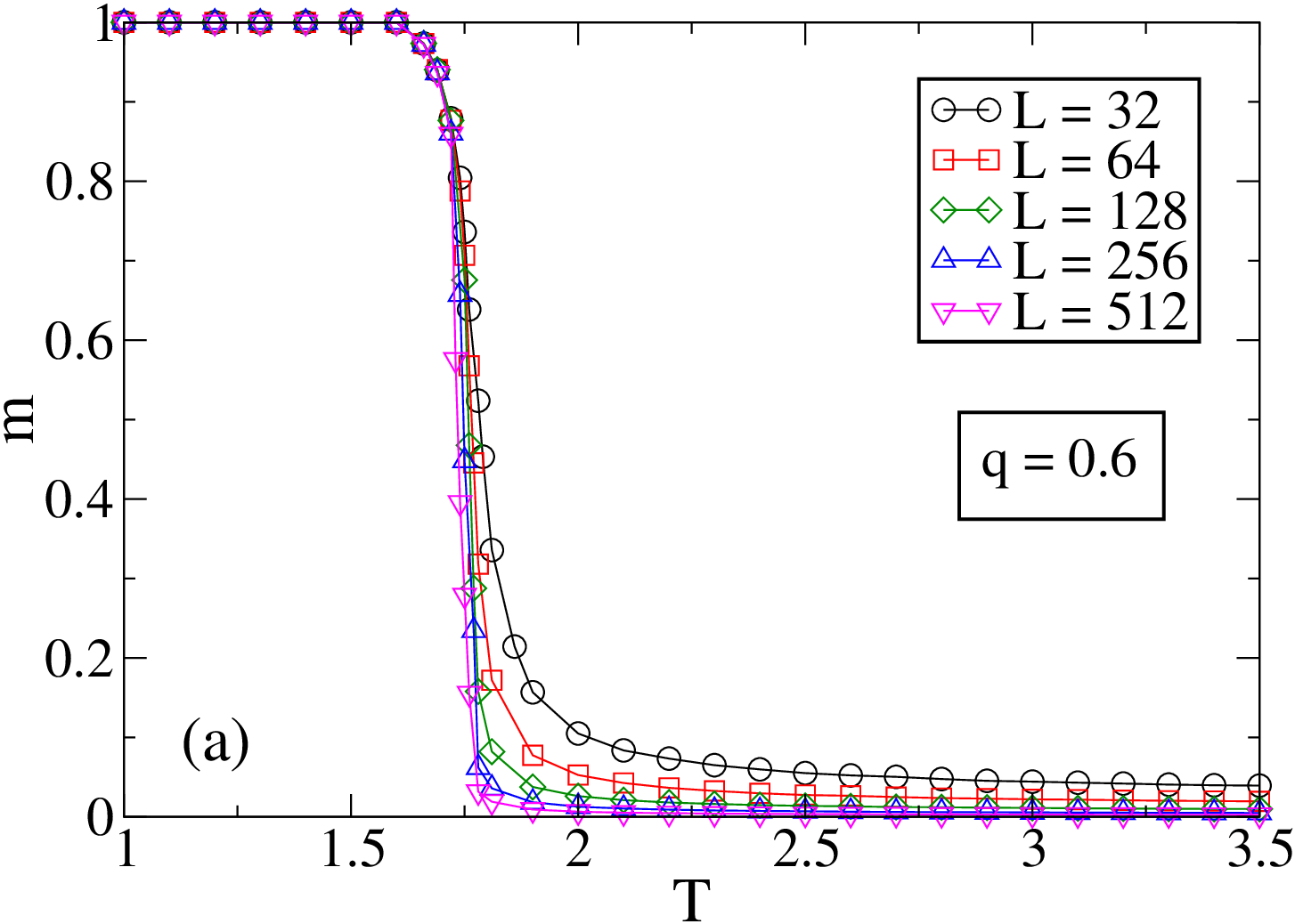} \hspace{1.5cm} %
\includegraphics[width=0.32\textwidth,angle=0]{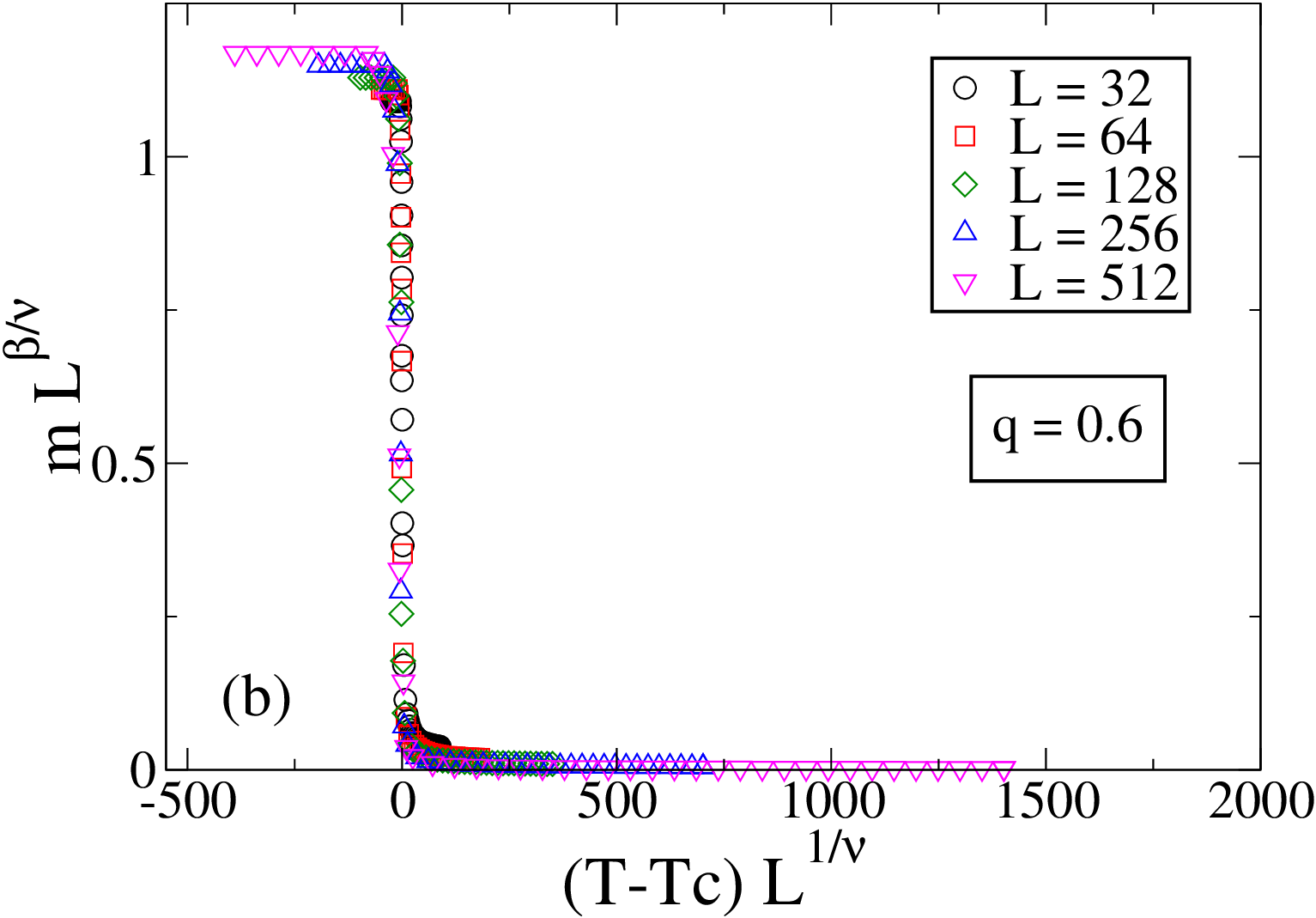} \\[0pt]
\vspace{1.0cm} \includegraphics[width=0.32\textwidth,angle=0]{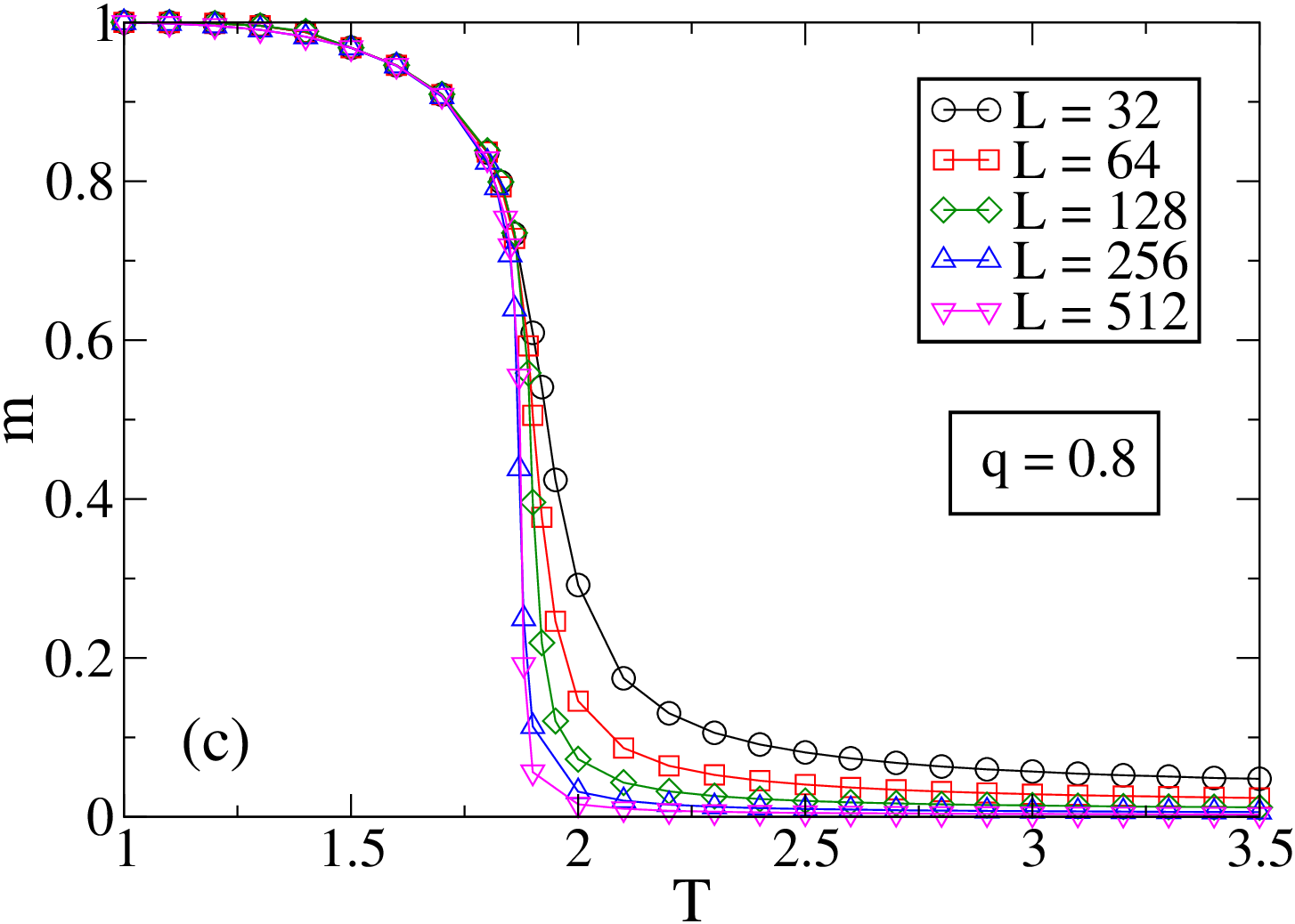} 
\hspace{1.5cm} \includegraphics[width=0.32\textwidth,angle=0]{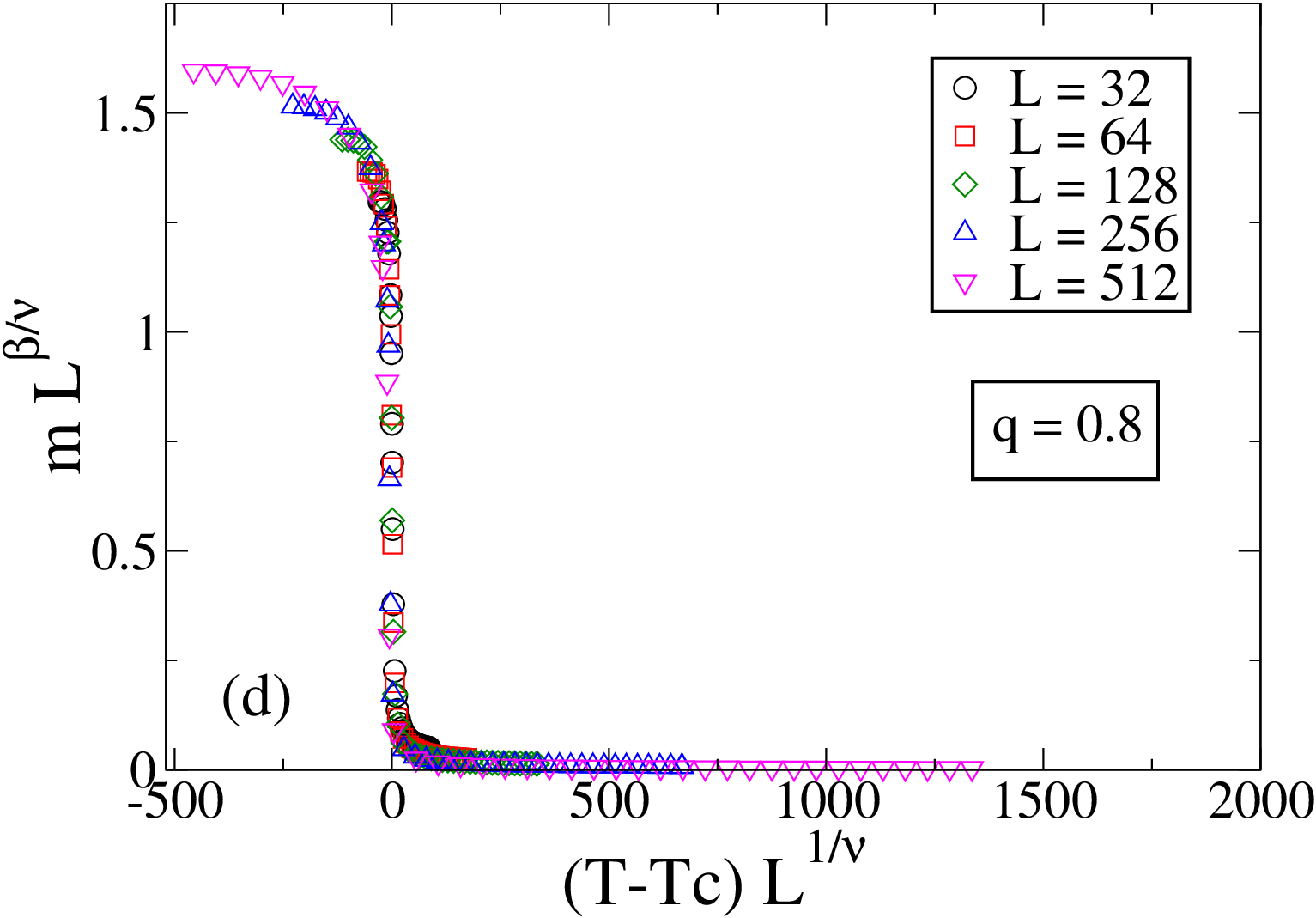} 
\\[0pt]
\vspace{1.0cm} \includegraphics[width=0.32\textwidth,angle=0]{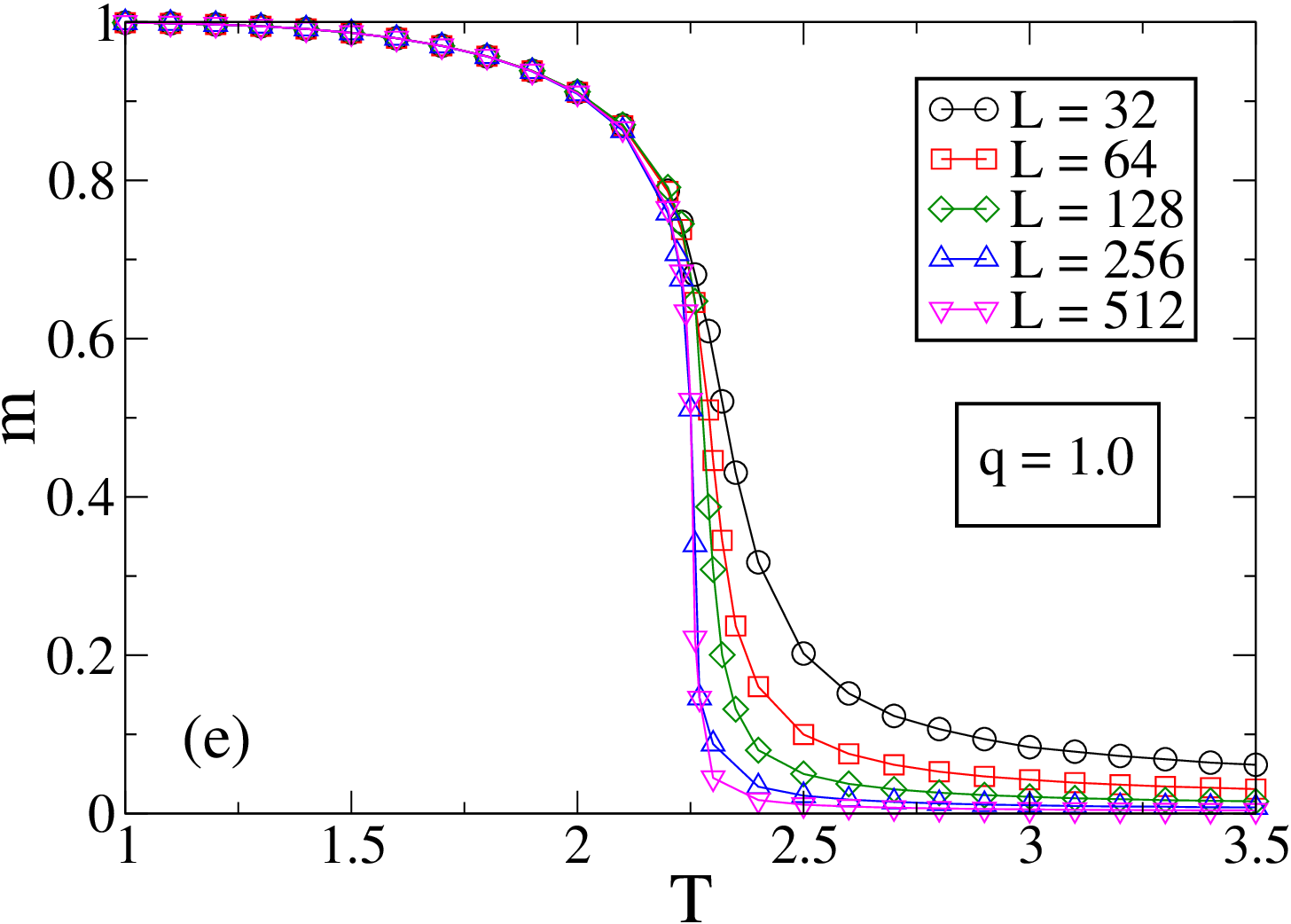} 
\hspace{1.5cm} \includegraphics[width=0.32\textwidth,angle=0]{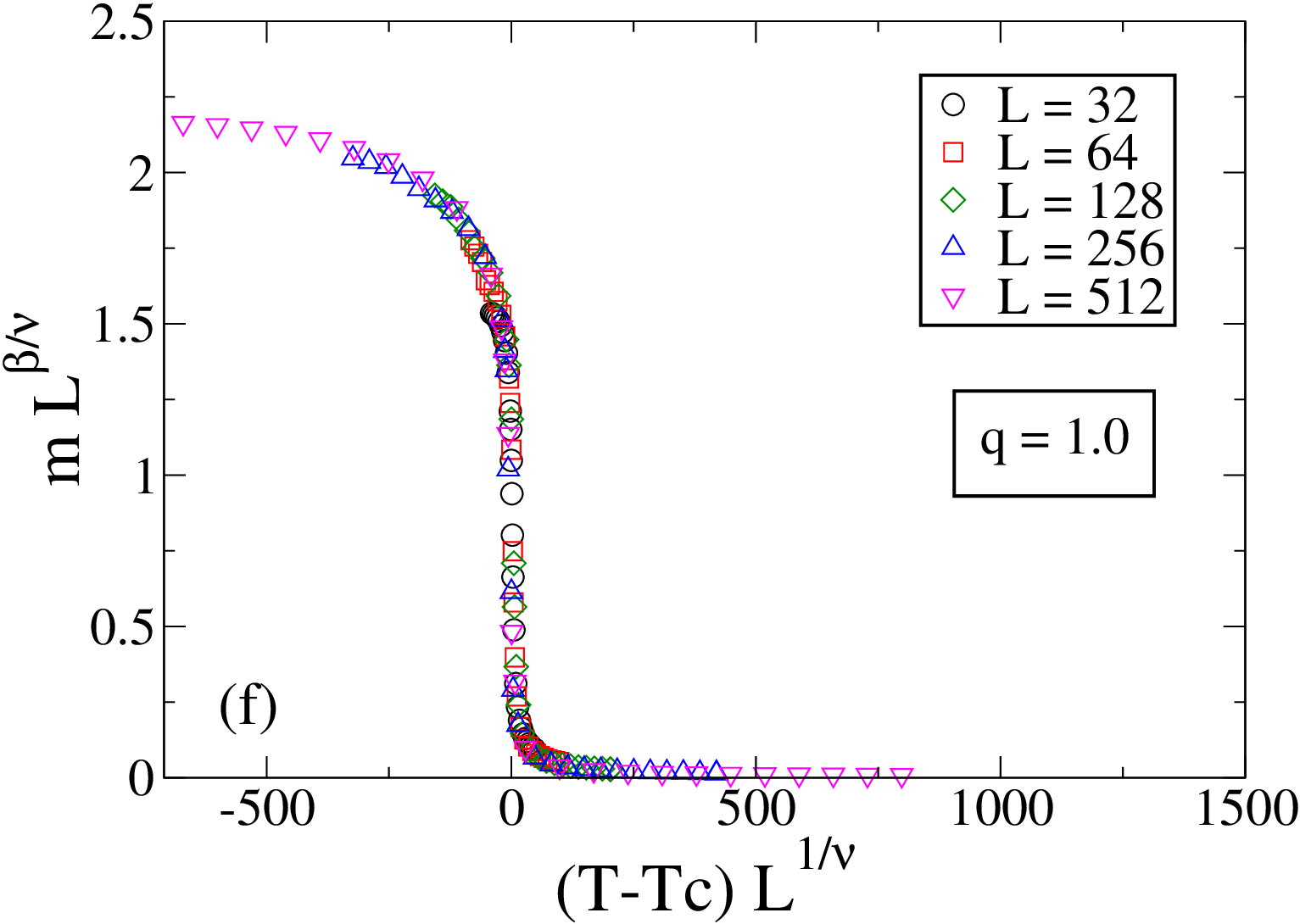}
\end{center}
\par
\caption{(Color online) Magnetization versus temperature (left side) and
scaling plots (right side) for some values of $0.5<q\leq 1.0$. We observe
phase transitions for all values of $q$ in that range at different critical
temperatures and with distinct critical exponents, that are given in Table
I.}
\label{Fig1}
\end{figure}

\section{Numerical results and finite size scaling}

In the following we will show results for the magnetization per spin, $m$,
and for the susceptibility $\chi$ and the specific heat $C$, which can be
obtained of the simulations from the fluctuation-dissipation relations, 
\begin{eqnarray}
\chi  &=&\frac{\langle \,m^{2}\rangle -\langle m\rangle ^{2}}{T}~,
\label{eq.09} \\
C &=&\frac{\langle \,e^{2}\rangle -\langle e\rangle ^{2}}{T^{2}}~,
\end{eqnarray}%
\noindent where $\langle \;\rangle $ stands for MC averages and $e$ is the
energy per spin (we have considered $J=k=1$ for simplicity). In Fig.~\ref%
{Fig1} it is shown the simulations in the range $0.5<q\leq 1.0$, where we
present on the left side the magnetization versus the temperature for $q=0.6$%
, $q=0.8$, and $q=1.0$. As may be observed on this figure, the magnetization
curves changes continuously from a ordered ferromagnetic phase to a
disordered paramagnetic one. Thus, the critical exponents and the critical
temperatures of the model can be obtained by the standard finite-size
scaling (FSS) forms, 
\begin{eqnarray}
m(T,L) &=&L^{-\beta /\nu }\;\tilde{m}((T-T_{c})\;L^{1/\nu })~,  \nonumber
\label{eq.10a} \\
\chi (T,L) &=&L^{\gamma /\nu }\;\tilde{\chi}((T-T_{c})\;L^{1/\nu })~, 
\nonumber \\
C(T,L) &=&L^{\alpha /\nu }\;\tilde{C}((T-T_{c})\;L^{1/\nu })~,  \nonumber \\
T_{c}(L) &=&T_{c}+a\;L^{-1/\nu }~,  \nonumber \\
&&
\end{eqnarray}

\begin{figure}[t]
\begin{center}
\includegraphics[width=0.35\textwidth,angle=0]{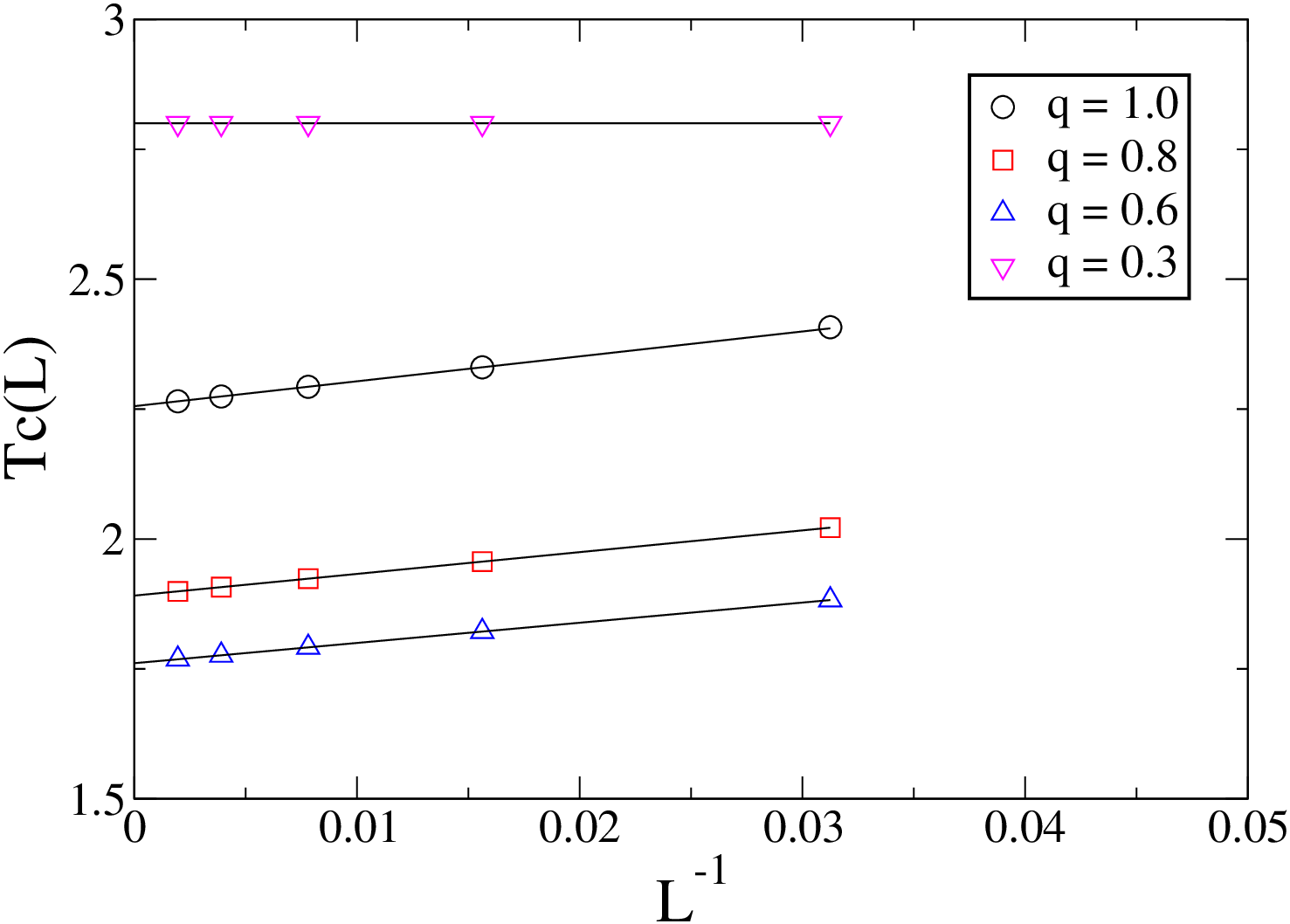}
\end{center}
\caption{(Color online) The pseudo-critical temperatures $T_{c}(L)$ versus $%
L^{-1}$ for some values of $q$. The extrapolation given us the critical
temperatures $T_{c}$ in the thermodynamic limit ($L^{-1}\to 0$). Notice
that for the $q\leq 0.5$ case the $T_{c}(L)$ values do not depend on $L$.}
\label{Fig_nova}
\end{figure}

\begin{table*}[tbp]
\begin{center}
\renewcommand\arraystretch{1.5} 
\begin{tabular}{|c|c|c|c|c|c|}
\hline
$q$ & $T_{c}$ & $\alpha$ & $\beta$ & $\gamma$ & $\nu$ \\ \hline
0.6 & 1.761\,$\pm$\,0.003 & 0.34\,$\pm$\,0.01 & 0.025\,$\pm$\,0.001 & 1.69\,$%
\pm$\,0.04 & 1.00\,$\pm$\,0.01 \\ 
0.8 & 1.891\,$\pm$\,0.007 & 0.15\,$\pm$\,0.02 & 0.075\,$\pm$\,0.002 & 1.71\,$%
\pm$\,0.04 & 1.00\,$\pm$\,0.01 \\ 
1.0 & 2.259\,$\pm$\,0.011 & 0.00\,$\pm$\,0.00 & 0.124\,$\pm$\,0.006 & 1.75\,$%
\pm$\,0.01 & 1.00\,$\pm$\,0.02 \\ \hline
\end{tabular}%
\end{center}
\caption{Three different entropic indexes in the range $0.5 < q \leq 1.0$
and its respective critical temperatures and exponents $\protect\alpha$, $%
\protect\beta$, $\protect\gamma$ and $\protect\nu$. For $q=1$, the critical
exponents and temperature are very close those expected, $\protect\alpha=0.0$%
, $\protect\beta = 0.125$, $\protect\nu = 1.0$, $\protect\gamma=1.75$ and $%
T_{c} = 2.269$. Notice that the exponents $\protect\nu$ are essentially the
same for the three cases, whereas $\protect\alpha$, $\protect\beta$, $%
\protect\gamma$ and $T_{c}$ depends on $q$. We have found a logarithmic
dependence of $\protect\alpha$ on the lattice size $L$ in the $q=1$ case, as
expected, which give us $\protect\alpha(q=1)=0$. The errors in the numerical
estimates of the critical temperatures and the critical exponents are also
presented.}
\label{Tab1}
\end{table*}

\begin{figure}[tbh]
\begin{center}
\vspace{0.5cm} \includegraphics[width=0.32\textwidth,angle=0]{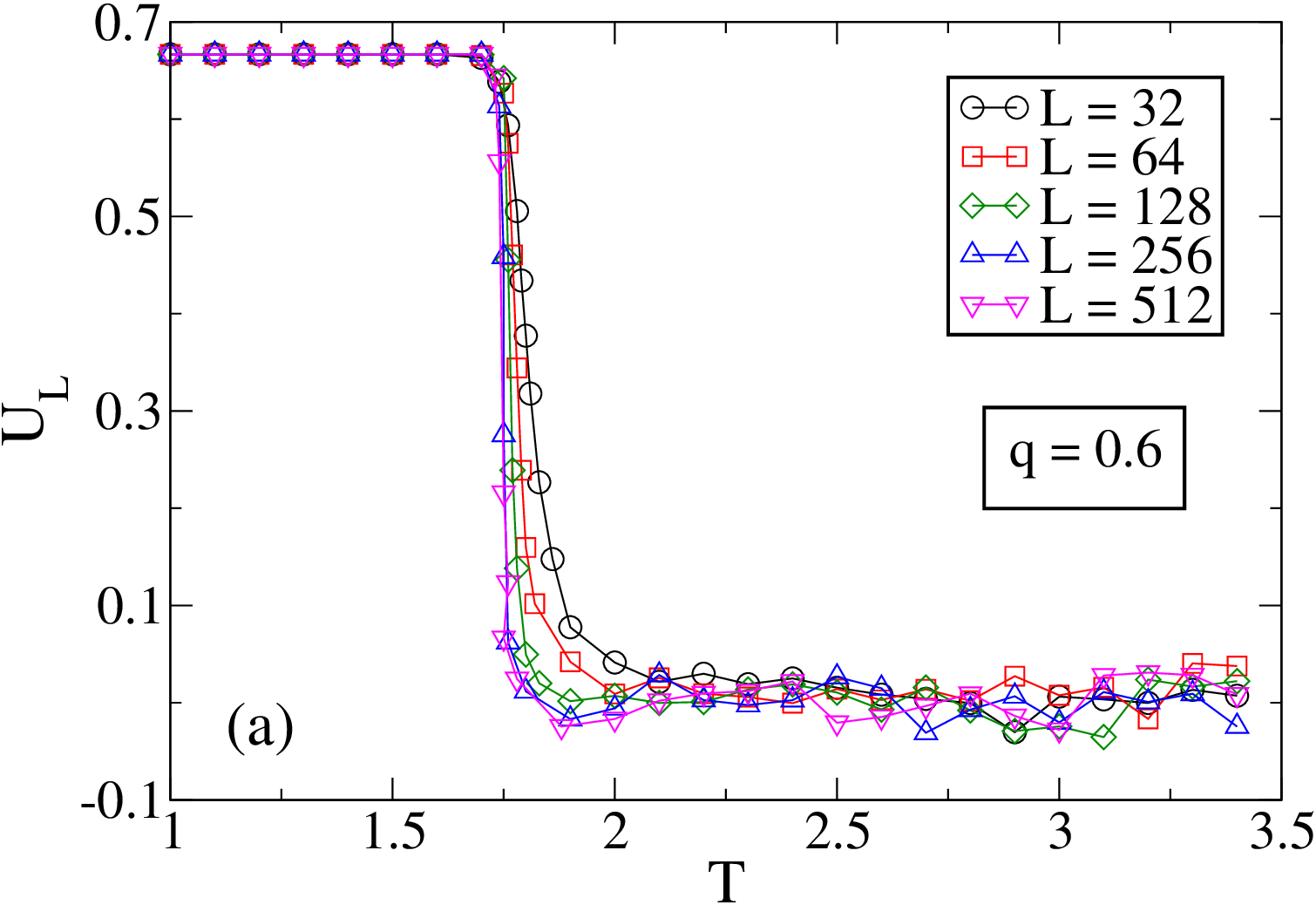}  
\hspace{1.5cm} \includegraphics[width=0.32\textwidth,angle=0]{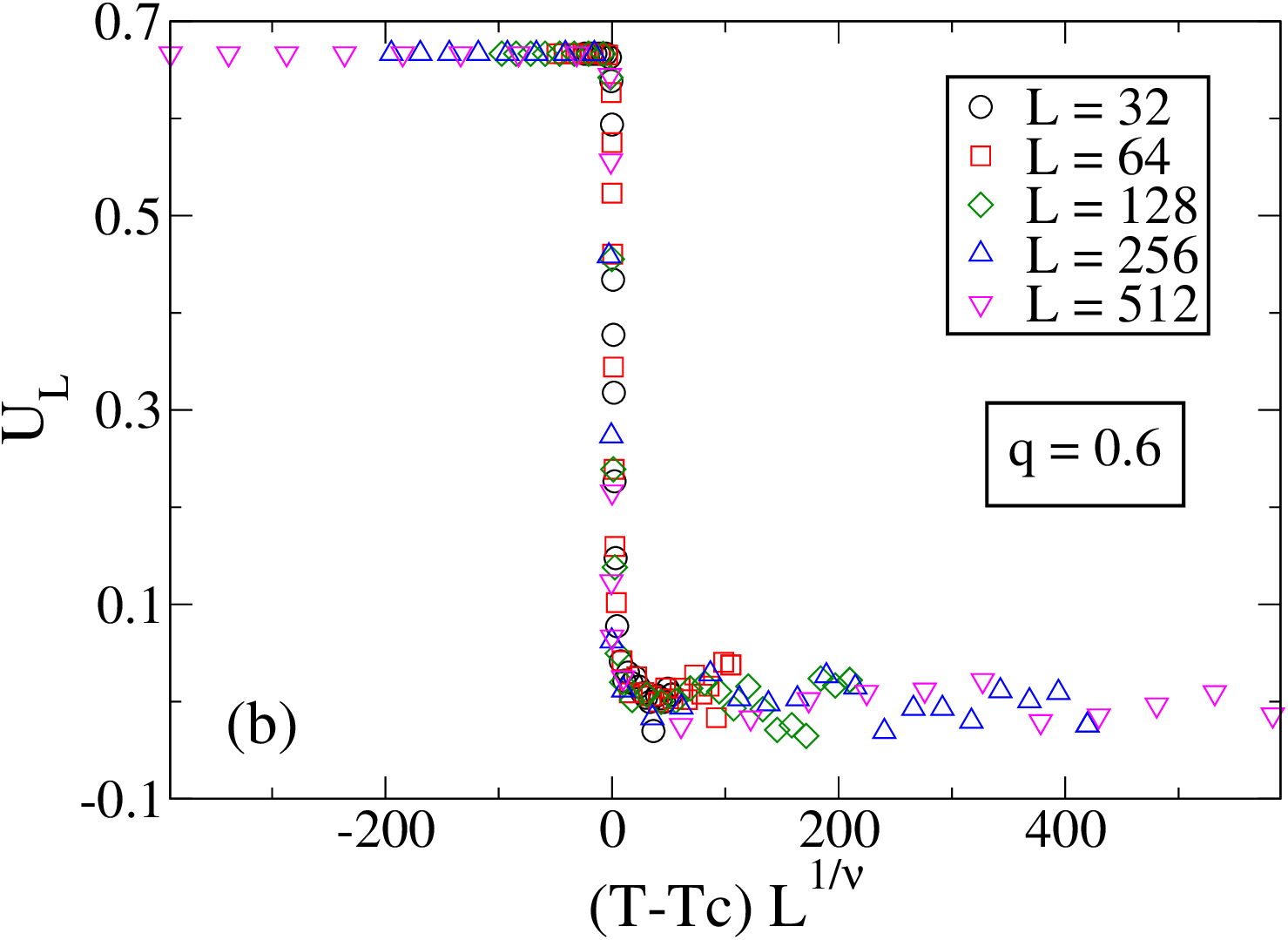}  
\\[0pt]
\vspace{1.0cm} \includegraphics[width=0.32\textwidth,angle=0]{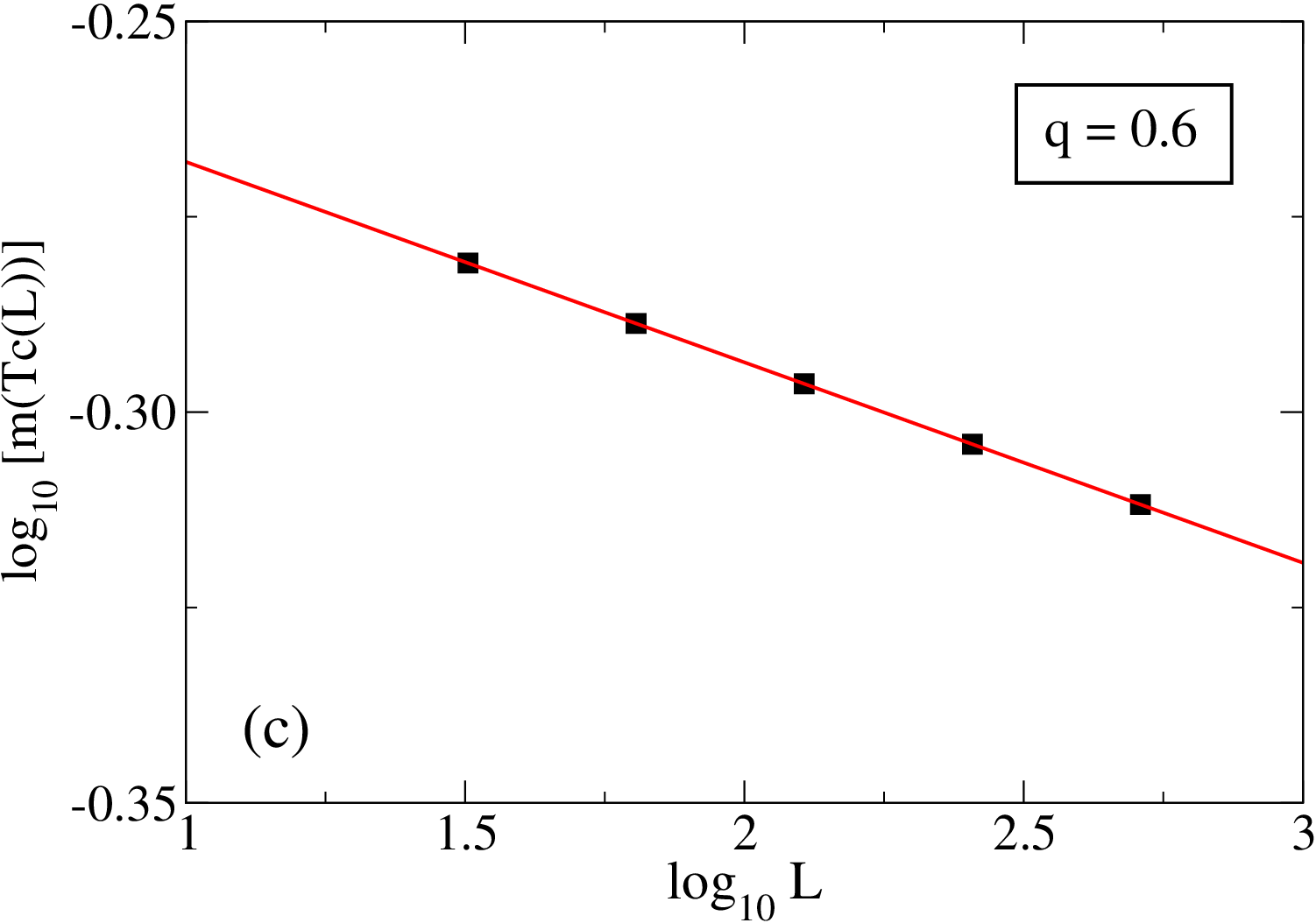}
\hspace{1.5cm} \includegraphics[width=0.32\textwidth,angle=0]{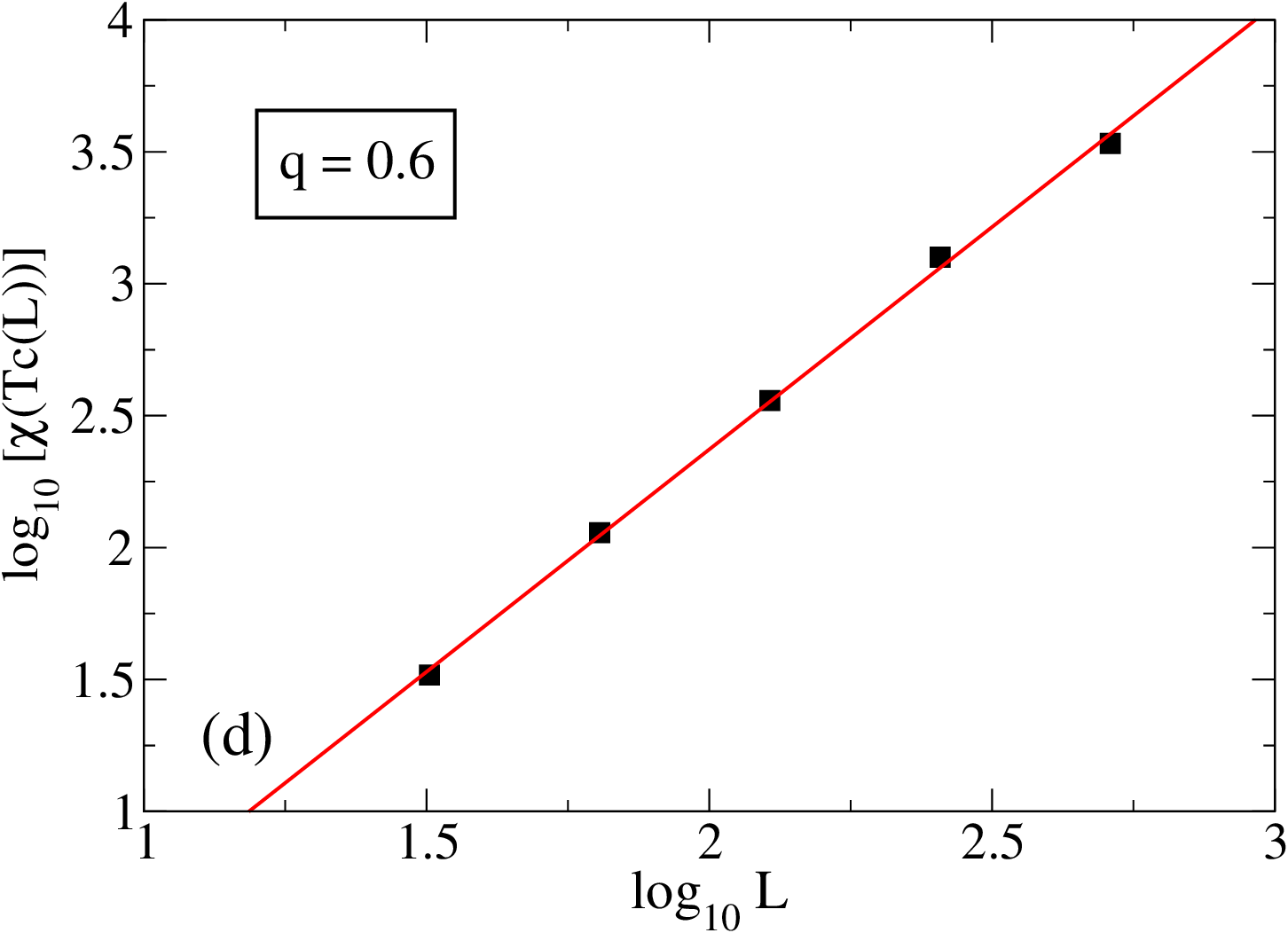}
\end{center}
\caption{(Color online) Upper figures: Binder cumulant, Eq.(\protect\ref%
{eq.10}), versus temperature for $q=0.6$ [Fig.~3(a)] and the best collapse
of data [Fig.~3(b)], based on the FSS in Eq.(\protect\ref{eq.11}). The
parameters are $\protect\nu=1.0$ and $T_{c}=1.761$. Lower figures:
magnetization values at each pseudo-critical temperature $T_{c}(L)$ for
various lattice sizes $L$ [Fig.~3(c)] and the corresponding values of the
susceptibility peaks [Fig.~3(d)], for $q=0.6$. The fittings in the log-log
scale give us the corresponding values of the critical exponents ratios $%
\protect\beta/\protect\nu$ and $\protect\gamma/\protect\nu$, $0.025 \pm
0.001 $ and $1.69 \pm 0.04$, respectively.}
\label{Fig2}
\end{figure}

\noindent where $a$ is a constant. In Eqs.~(\ref{eq.10a}), the exponent $%
\beta $ is related to the behavior of the magnetization near the critical
point $T_{c}$, $\nu $ is related to the divergence of the correlation
length, $\gamma $ governs the divergence of the susceptibility at the
critical point, $\alpha $ is related to the divergence of the specific heat
at $T_{c}$ and $\tilde{m}$, $\tilde{\chi}$ and $\tilde{C}$ are scaling
functions. The critical temperatures of the infinite lattices $T_{c}$ were
obtained by extrapolating the $T_{c}(L)$ values given by the susceptibility
peaks positions$^{1}$\footnotetext[1]{%
Equivalently, we can determine the $T_{c}(L)$ values by the maxima of the
specific heat curves.} (see Fig. \ref{Fig_nova}). On the other hand, the
exponents $\nu $ were obtained by means of the Binder cumulant \cite{binder}%
, defined as 
\begin{equation}
U_{L}=1-\frac{\langle \,m^{4}\rangle }{3\langle \,m^{2}\rangle ^{2}}~,
\label{eq.10}
\end{equation}%
\noindent which has the FSS form 
\begin{equation}
U_{L}=\tilde{U_{L}}((T-T_{c})\;L^{1/\nu })~,  \label{eq.11}
\end{equation}%
\noindent where $\tilde{U_{L}}$ is a scaling function that is independent of 
$L$. The error bars in the estimations of $\nu$ were obtained following the
standard procedure for collapsing data of the Binder cumulant in the
finite-size scaling approach, i.e., by monitoring small variations around
the best collapsing pictures. In Fig.\ref{Fig2} we show, as an example, the
Binder cumulant for $q=0.6$ [Fig.~3(a)] and the best collapse
of data [Fig.~3(b)], based on Eq.(\ref{eq.11}), obtained with
the critical temperature $T_{c}$ and the exponent $\nu $ given in Table \ref%
{Tab1}. Also in Fig.\ref{Fig2} we show, for $q=0.6$, the values of the
magnetization at the pseudo-critical points $T_{c}(L)$ for various lattice
sizes $L$ [Fig.~3(c)] and the corresponding values of the susceptibility
peaks positions [Fig.~3(d)], in the log-log scale. Linear fitting of 
data yield the parameters: 
\begin{eqnarray}
\beta /\nu  &=&0.025\pm 0.001~,  \label{eq.11a} \\
\gamma /\nu  &=&1.69\pm 0.04~.
\end{eqnarray}%
\noindent By repeating the fitting procedures of the specific heat peaks
positions versus the lattice size $L$, in the log-log scale, we have
estimated the ratio $\alpha /\nu =0.34\pm 0.01$, for $q=0.6$ (see Table \ref%
{Tab1}). We have calculated the exponent $\nu $ by means of the Binder
cumulant, as above-discussed, which allowed us to estimate the critical
exponents $\alpha $, $\beta $ and $\gamma $. The procedure was the same for
the other values of the entropic index $q$, and the best collapse of the
magnetization data, presented on the right side of Fig.~\ref{Fig1}, supports
the validity of the FSS forms in Eqs.~(\ref{eq.10a}) and the reliability of
the numerical results for the critical exponents. The obtained numerical
results are summarized on Table \ref{Tab1}. Note that for $q=1$, the
critical exponents $\alpha $, $\beta $, $\gamma $ and $\nu $ are quite close
to the exact known values of the standard 2D Ising model, as expected.
However, for different values of $q$ in this range, we have found different
values of the exponents $\alpha $, $\beta $ and $\gamma $, as we can see in
Table \ref{Tab1}, whereas the values of $\nu $ are the same, within the
determined uncertainty. These results will be discussed in with more details
bellow.

\begin{figure}[tbh]
\begin{center}
\vspace{0.5cm} \includegraphics[width=0.32\textwidth,angle=0]{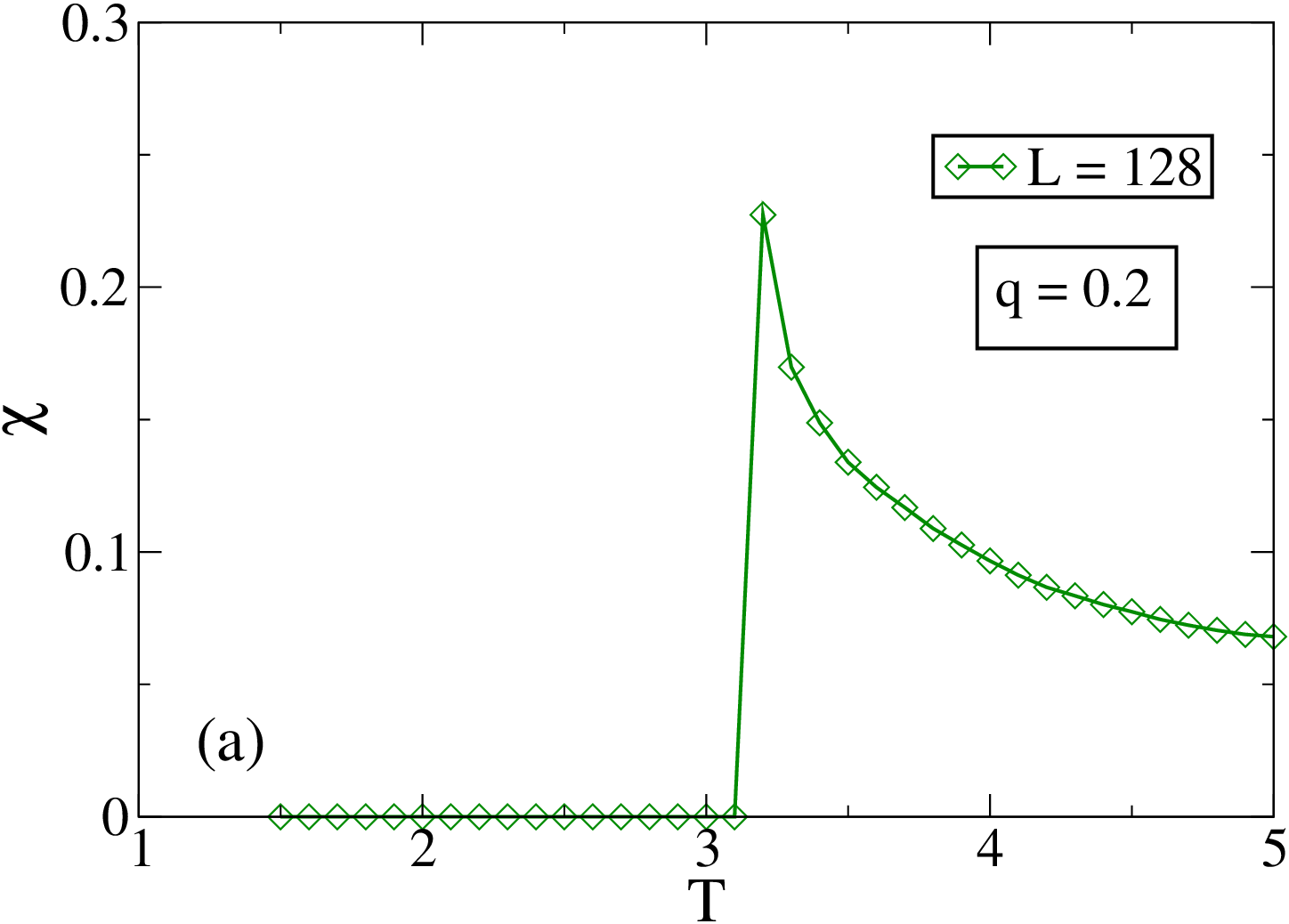}
\hspace{1.5cm} \includegraphics[width=0.32\textwidth,angle=0]{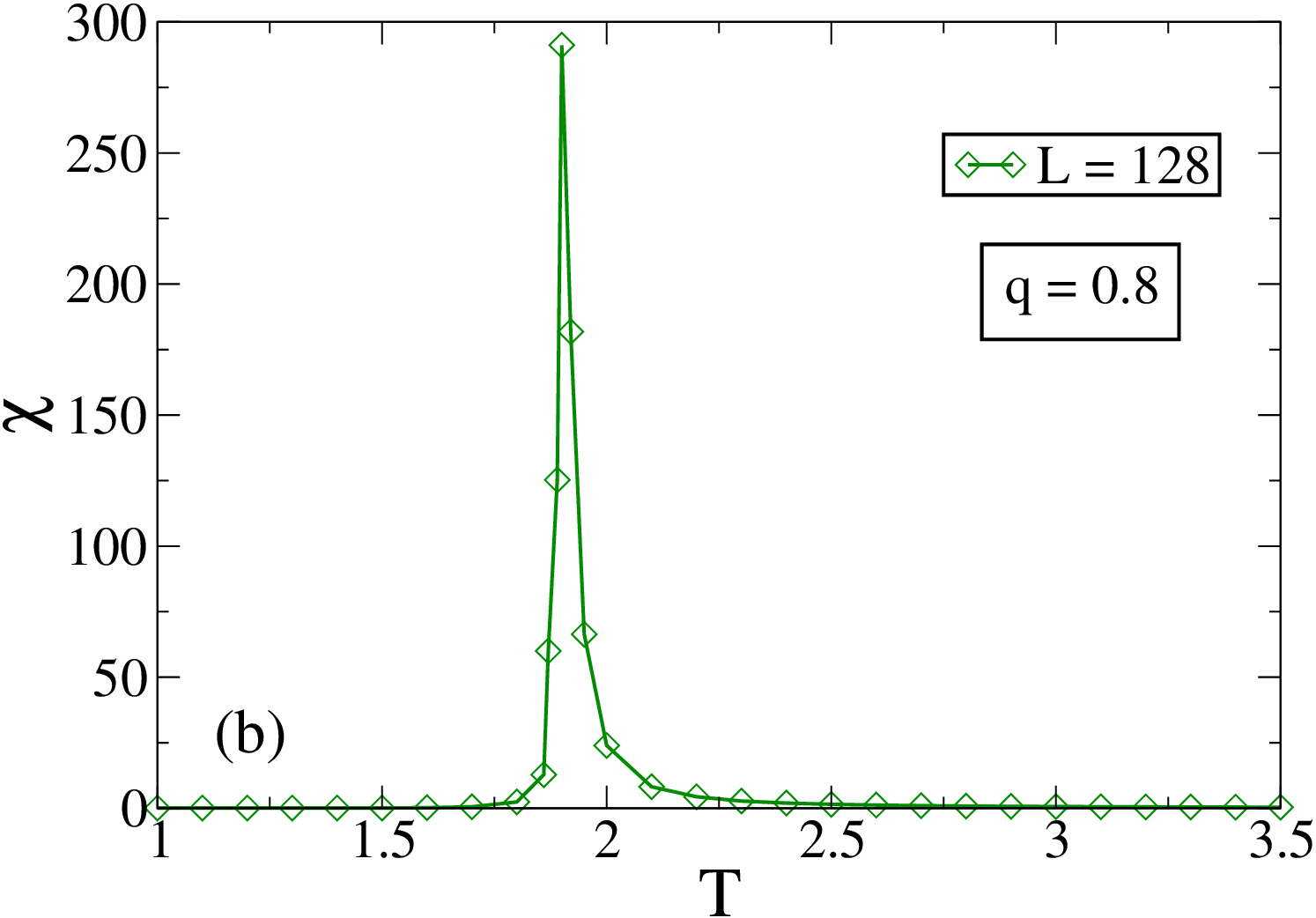}
\\[0pt]
\vspace{1.0cm} \includegraphics[width=0.32\textwidth,angle=0]{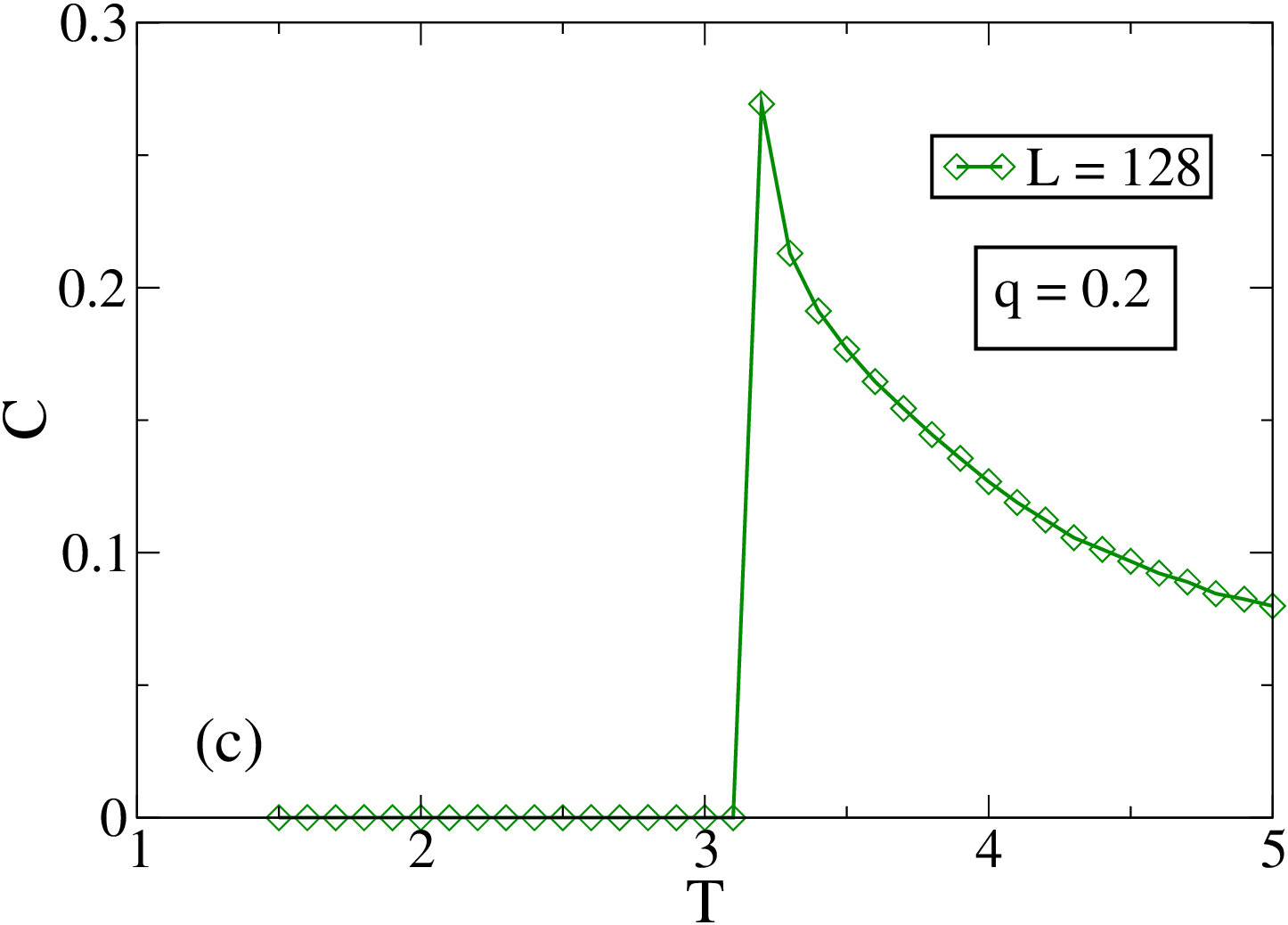} 
\hspace{1.5cm} \includegraphics[width=0.32\textwidth,angle=0]{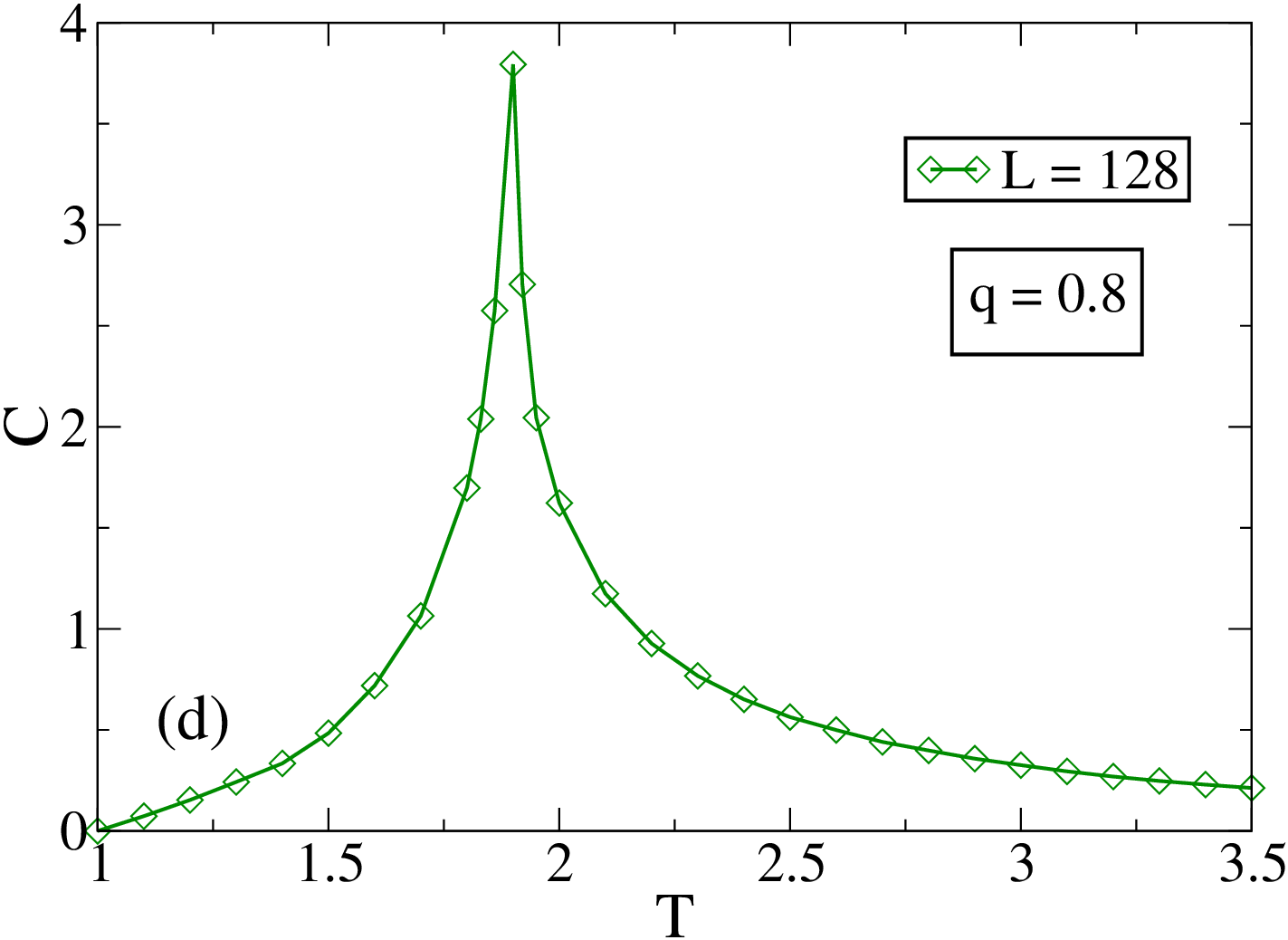} 
\\[0pt]
\vspace{1.0cm} \includegraphics[width=0.32\textwidth,angle=0]{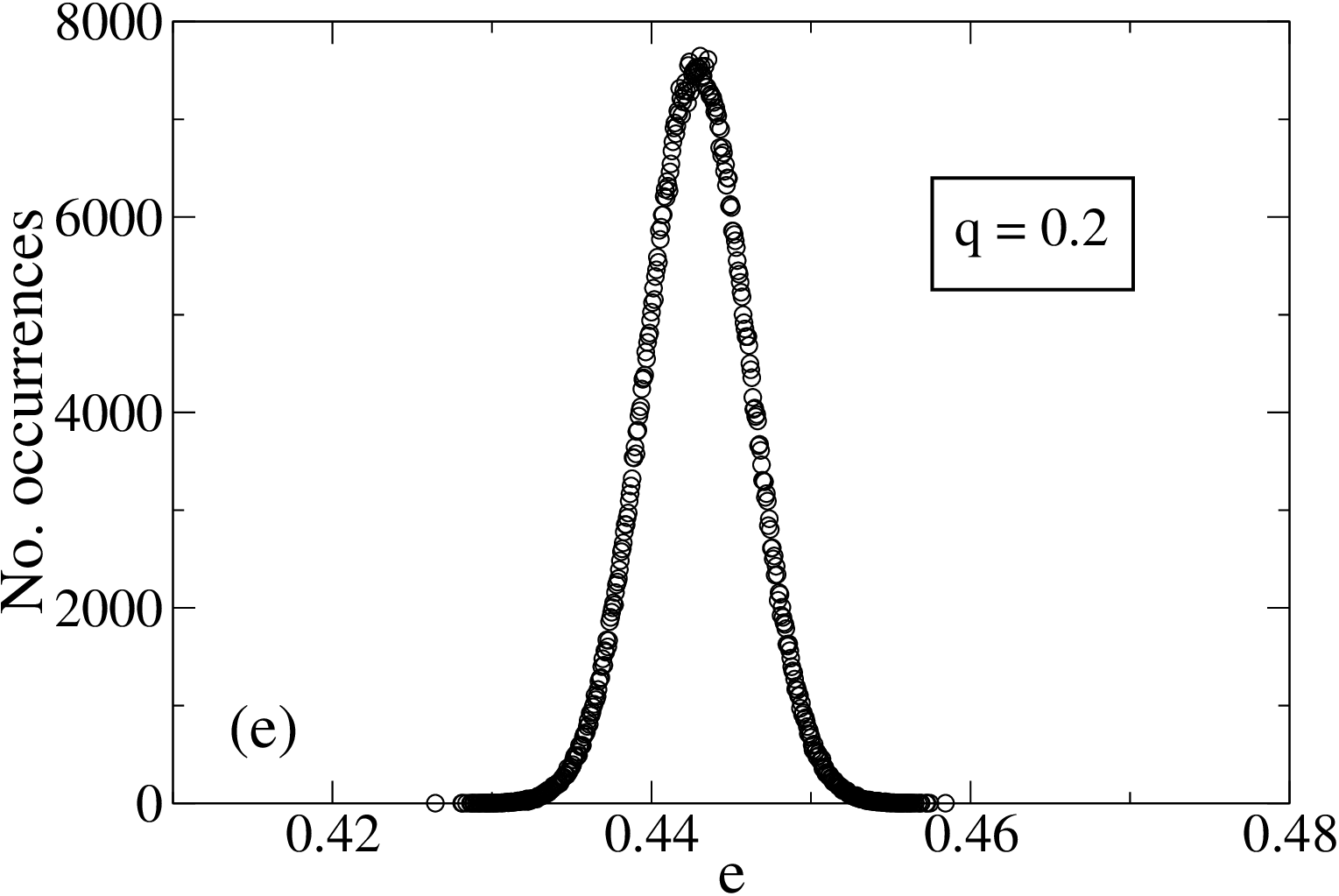} 
\hspace{1.5cm} \includegraphics[width=0.32\textwidth,angle=0]{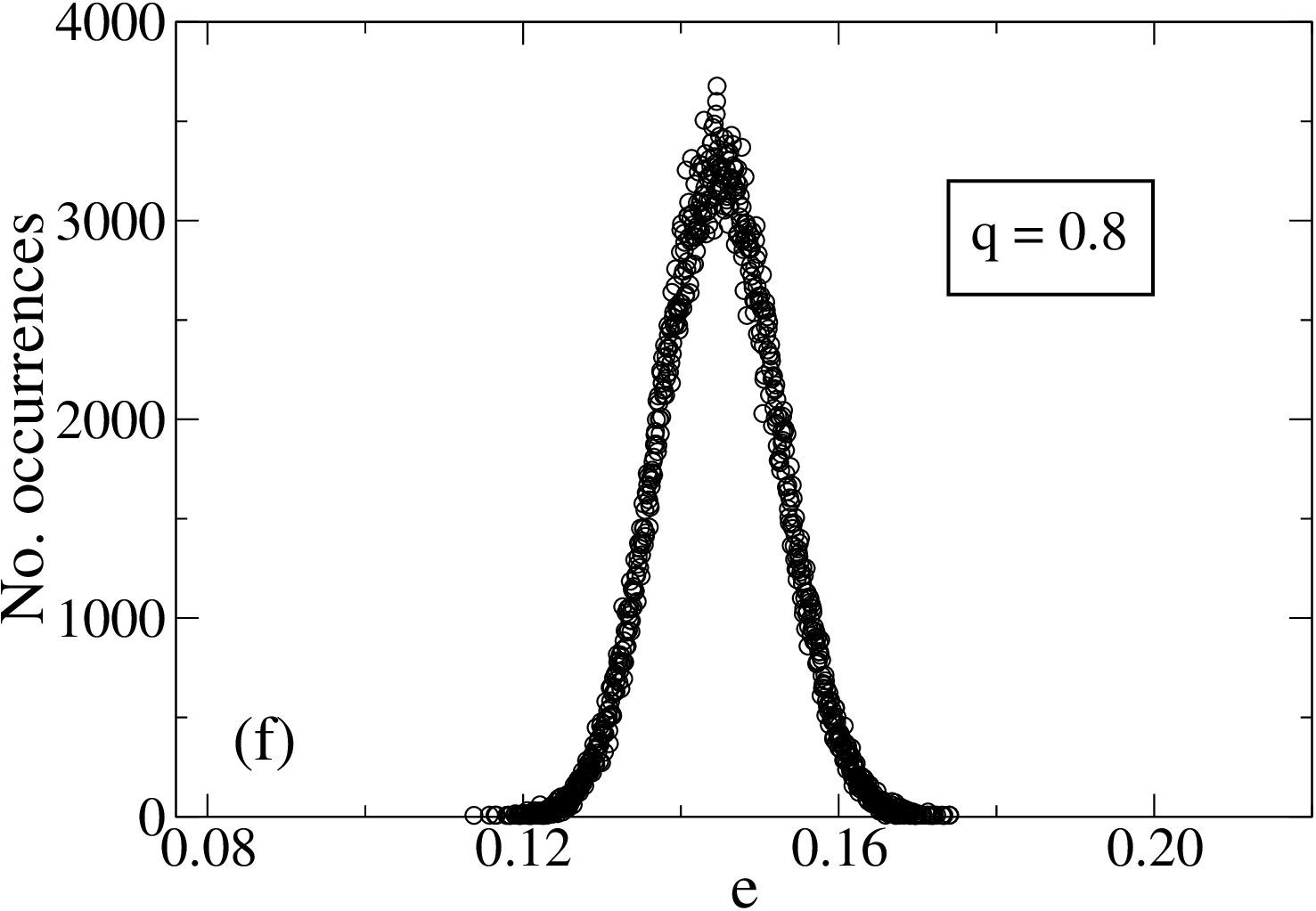}
\end{center}
\caption{(Color online) Results for the magnetic susceptibility $\protect%
\chi $ for $L=128$ and typical cases of $q<0.5$ [$q=0.2$, (a)] and $%
q>0.5$ [$q=0.8$, (b)]. It is also shown the specific heat curves for the
same values of $q$ [(c) and (d), respectively]. Although it is possible to observe a 
jump on the left figures, the histograms of the energy states visited during 
the MC simulation [Figs.~4(e) and 4(f)], at the corresponding critical temperatures, 
show only one-peak structures, indicating continuous phase transitions, even for the case $q<0.5
$. We have defined the energy as the fraction of unhappy bonds in the
system, $e=(E+2N)/(4N)$, where $E$ is the total energy given by Eq. (\protect
\ref{eq.08}) and $N$ is the total number of spins.}
\label{Fig3}
\end{figure}

In Fig.~\ref{Fig3} (right side) we show as example the susceptibility and
the specific heat for $q=0.8$, as well as a histogram of the energy states
visited during the dynamics of the system, at the critical temperature. This
histogram shows only one peak, i.e., we have a continuous phase transition,
as shown in the magnetization curves, Fig.\ref{Fig1}.

\begin{figure}[tbh]
\begin{center}
\vspace{0.5cm} \includegraphics[width=0.32\textwidth,angle=0]{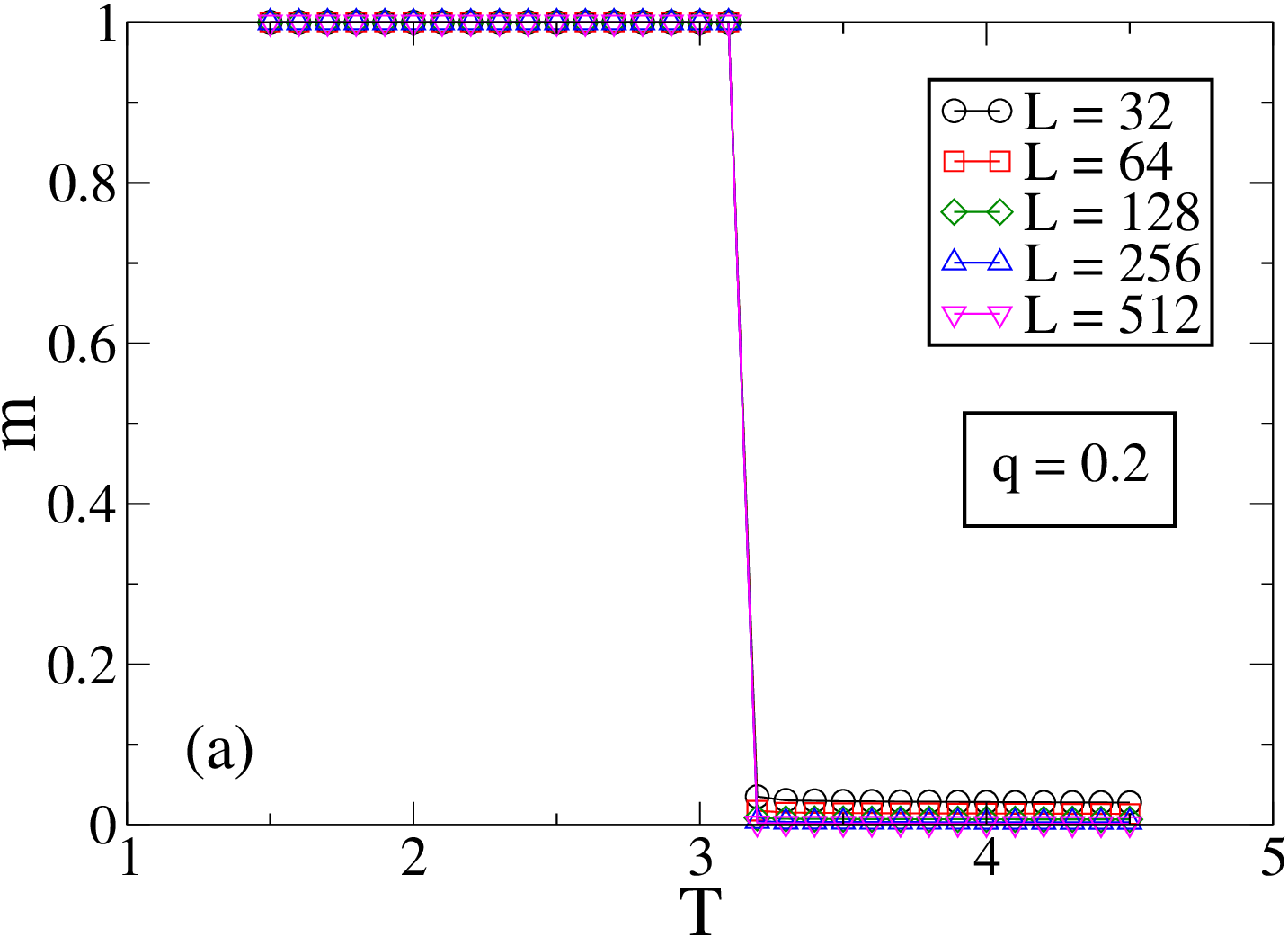} 
\hspace{1.5cm} \includegraphics[width=0.32\textwidth,angle=0]{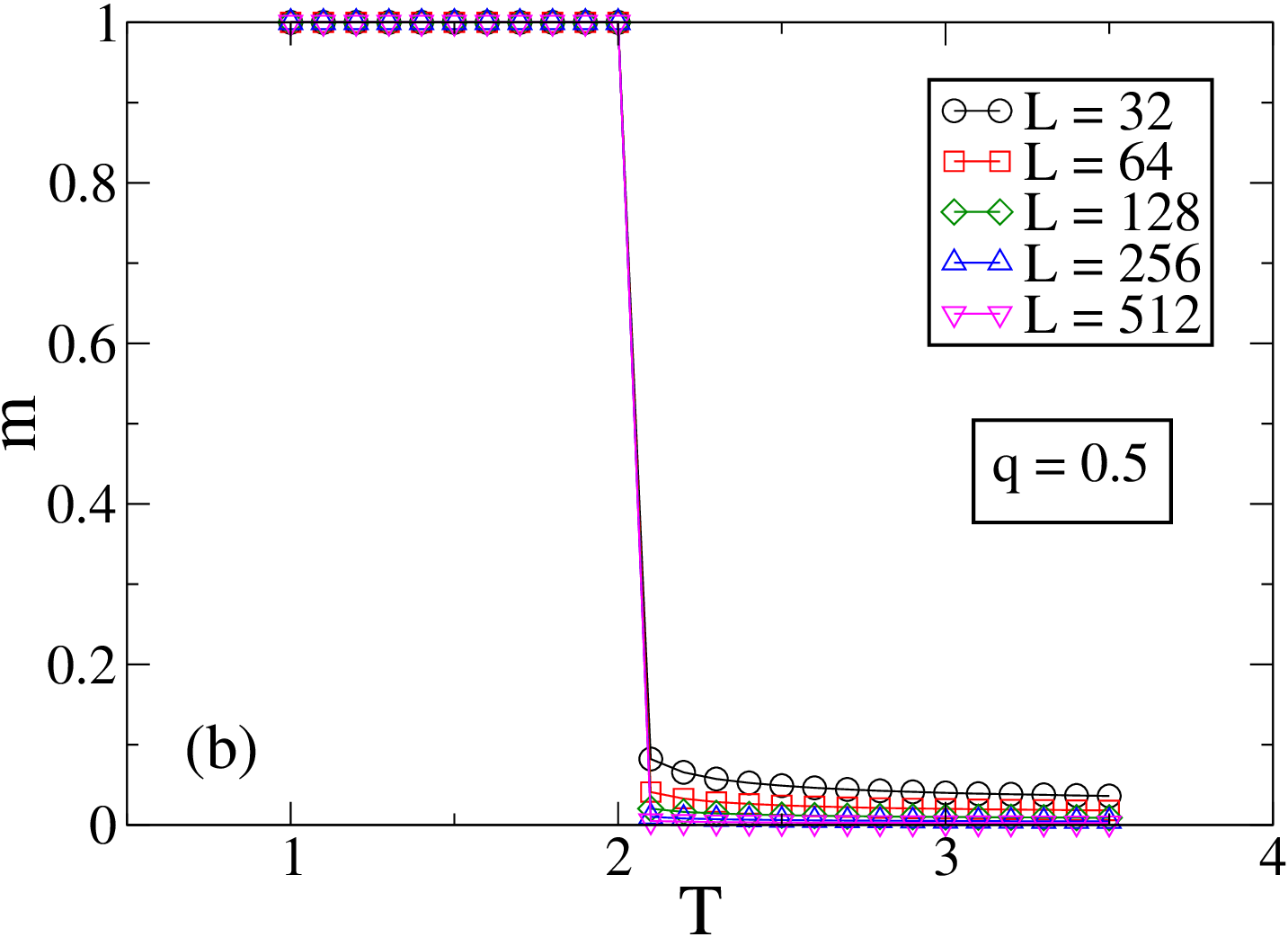}
\end{center}
\caption{(Color online) Magnetization versus temperature for $q = 0.2$ and $%
q = 0.5$. We observe jumps on the curves at the corresponding critical
points, but these $T_{c}$ values are independent of the lattice size,
showing that the scaling naturally occurs on the $q\leq 0.5$ regime.}
\label{Fig4}
\end{figure}

In Fig.~\ref{Fig4} it is shown the behavior of the magnetization as a
function of the temperature for two cases of $q\leq 0.5$. One can see that
the critical temperatures, $T_{c}$ $^{2}$\footnotetext[2]{%
According to Ref.\cite{2008_EPJB_62_337}, the critical temperatures in this
regime are given by $T_{c}=4~(1-q)$.}, are the same for all lattice sizes, 
as shown in Fig.~\ref{Fig_nova}. This result is a consequence of the cutoff of the escort
distribution, Eq.(\ref{eq.03}), and the magnetization jumps at $T_{c}$
from $m=1$ to $m=0$ (for more details, see \cite{2008_EPJB_62_337}). The cutoff also affects the
susceptibility and the specific heat, as we can see in Figs.~4(a) and
4(c), respectively. However, if we compute histograms of the energy states visited during the dynamics$^{3}$%
\footnotetext[3]{%
The energy per spin curves show jumps at the critical temperatures $T_{c}$
for the $q\leq 0.5$ cases, but the histograms of the energy states visited
clearly show one-peak structures, indicating the occurrence of continuous
phase transitions. In other words, the jumps are only an effect of the
cutoff of the Tsallis distribution, as in the magnetization curves.}, at the critical
temperatures, we can verify that we have only one peak for all $q\in \lbrack
0,1]$, indicating the occurrence of continuous phase
transitions [see Figs.~4(e) and (f)]. In other words, the cutoff keep
the MC simulation trapped in the ground state for $T < T_{c}$ and the
thermodynamic quantities suddenly change at $T_{c}$.


\begin{figure}[tbh]
\begin{center}
\vspace{0.5cm} \includegraphics[width=0.3\textwidth,angle=0]{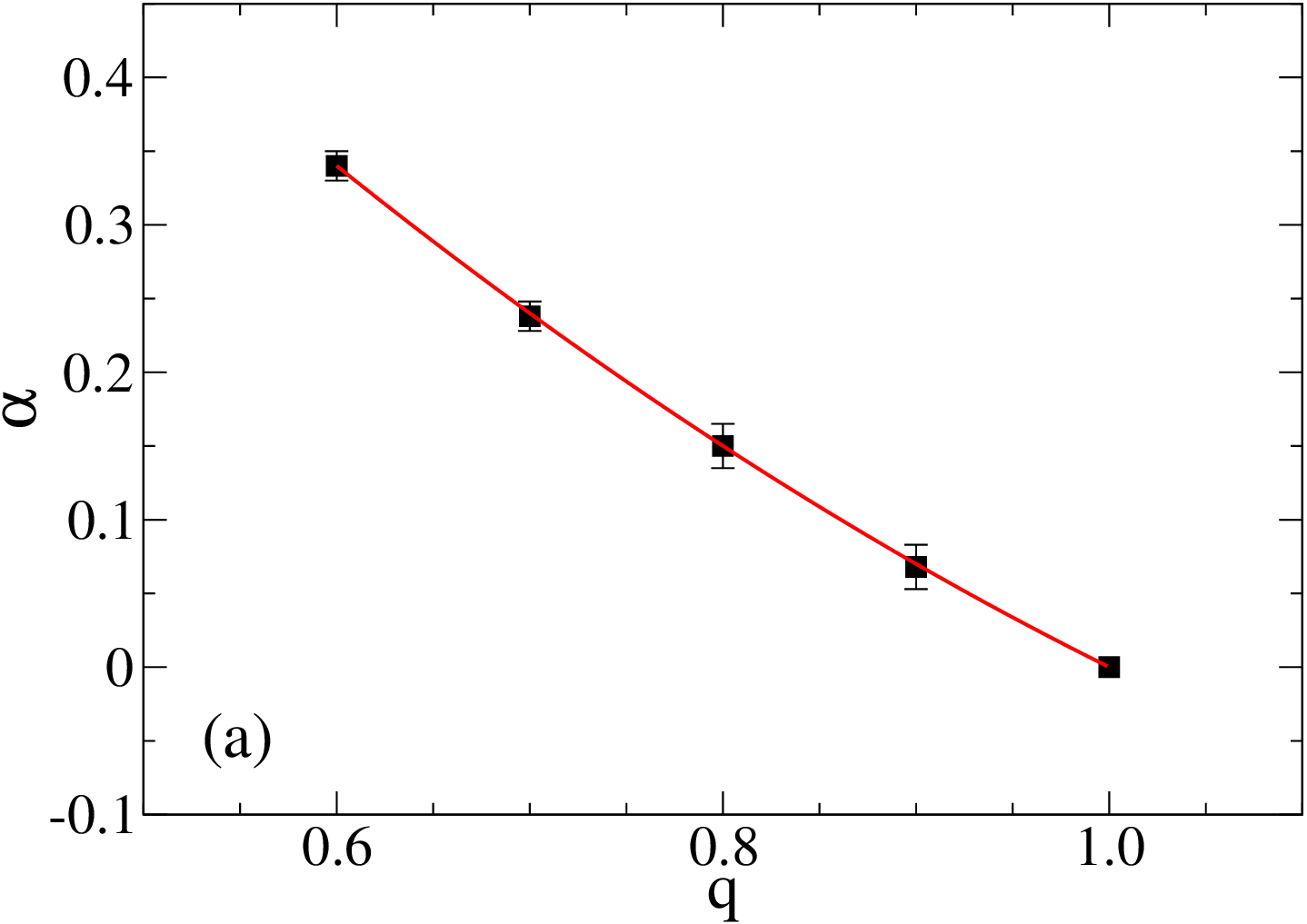} 
\hspace{1.5cm} \includegraphics[width=0.3\textwidth,angle=0]{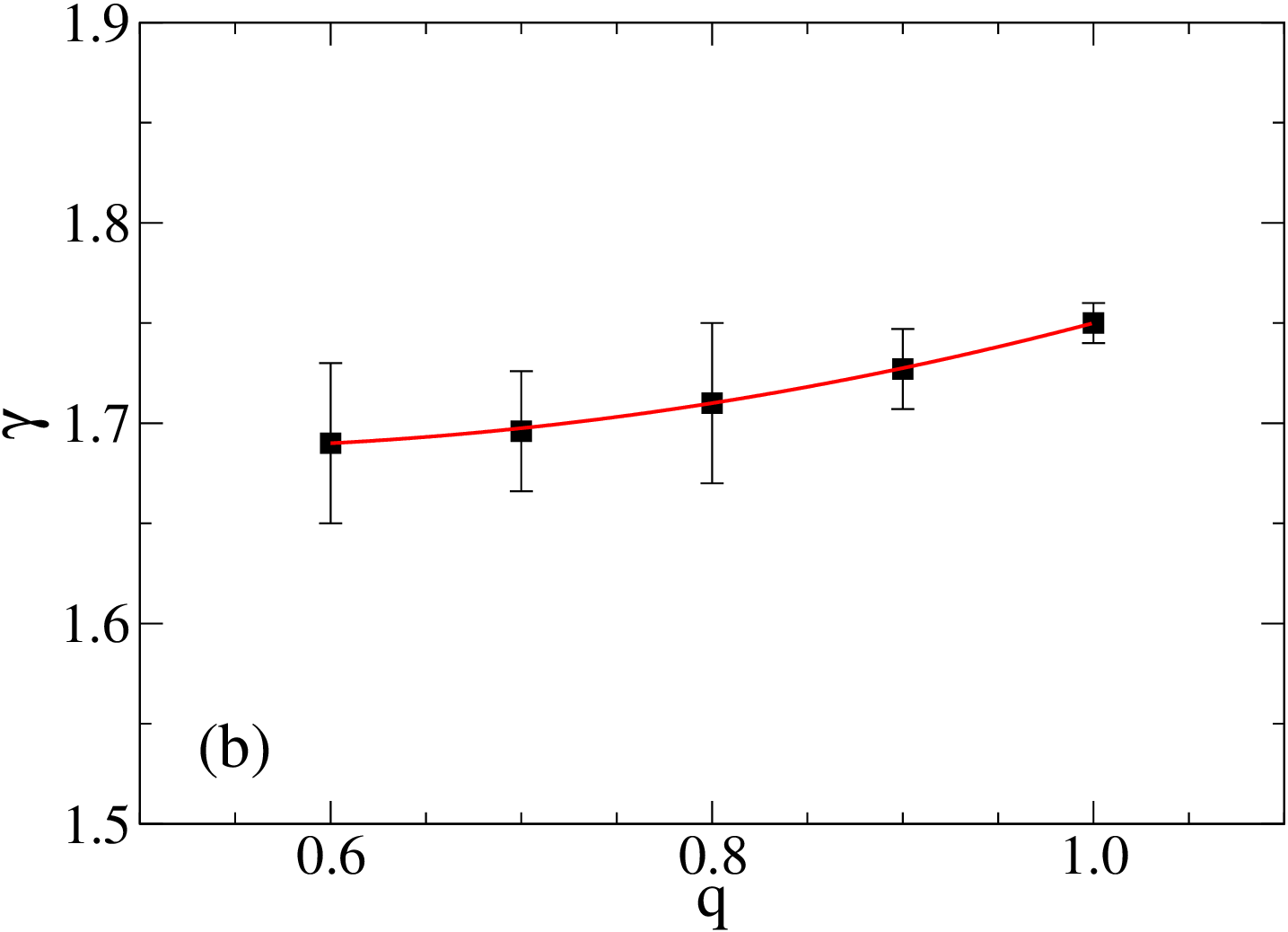} \\%
[0pt]
\vspace{1.0cm} \includegraphics[width=0.3\textwidth,angle=0]{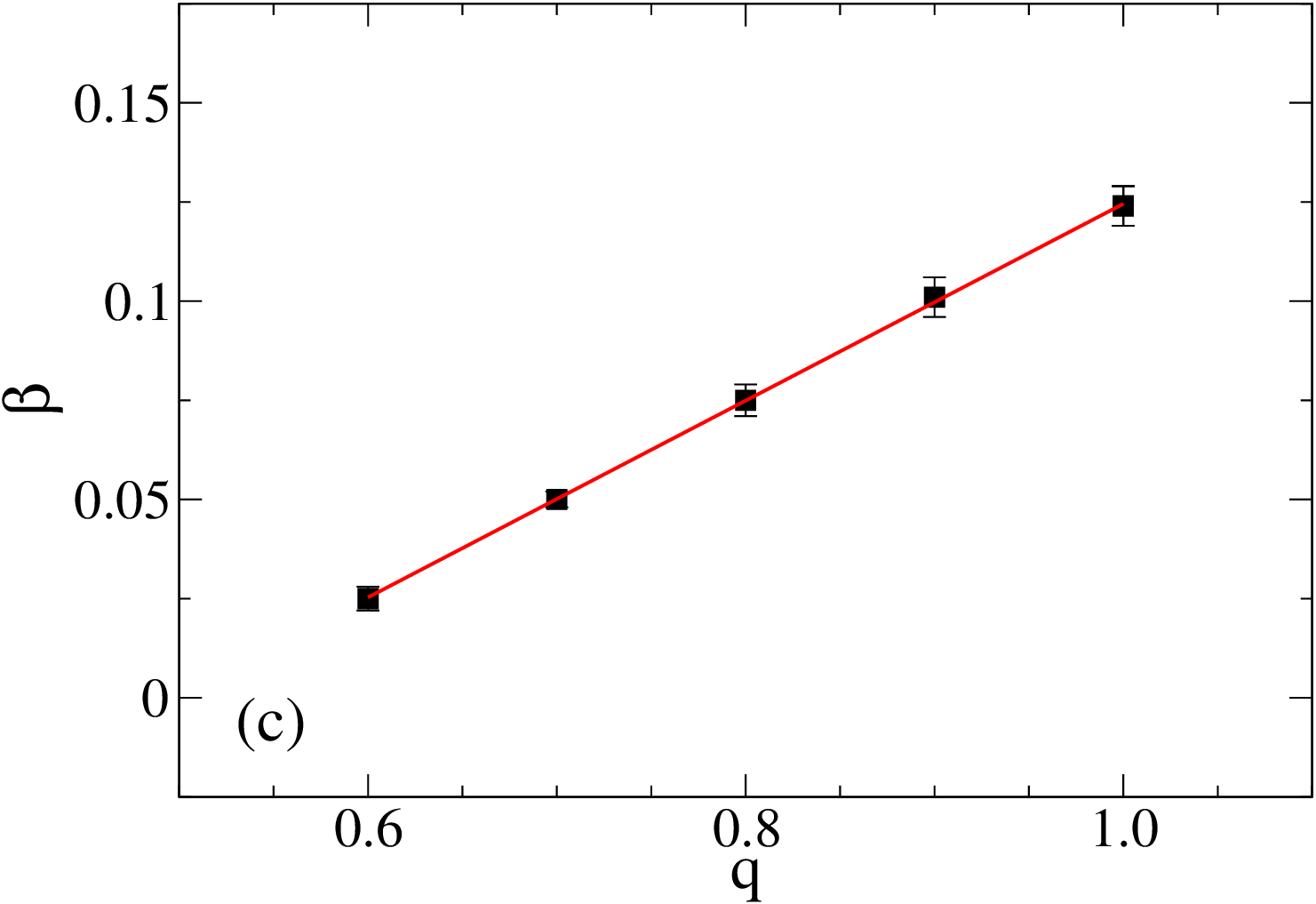}
\end{center}
\caption{(Color online) The critical exponents $\protect\alpha$, $\protect%
\beta$ and $\protect\gamma$ as functions of the entropic index $q$. In the
range $0.5< q\leq1.0$, the dependencies on $q$ are given by $\protect\alpha%
(q)=(10\,q^{2}-33\,q+23)/20$, $\protect\beta(q)= (2\,q-1)/8$ and $\protect%
\gamma(q)=(q^{2}-q+7)/4$. The error bars for $\protect\alpha$, $\protect%
\beta $ and $\protect\gamma$ fittings are less than 5\%.}
\label{Fig5}
\end{figure}


We can see from Table \ref{Tab1} that the critical exponent $\nu $ does not
depends on $q$ in the range $0.5<q\leq 1.0$, and we conjecture that the
correct value for any $q$ is $\nu =1.0$. However, $\alpha $, $\beta $ and $%
\gamma $ depend on the value of $q$. Fitting the numerical values of $\alpha 
$ with a second-order polynomial function of $q$, we have found that $\alpha
(q)=0.5\,q^{2}-1.65\,q+1.15$, for $0.5<q\leq 1.0$ (see Fig.~\ref{Fig5}), or 
\begin{equation}
\alpha (q)=\frac{1}{20}(10q^{2}-33q+23),  \label{eq.12a}
\end{equation}%
which give us the exact known value $\alpha (q=1)=0$ $^{4}$\footnotetext[4]{%
We have found a logarithm dependence of $\alpha $ on the lattice size $L$ in
the $q=1$ case, as expected, which give us $\alpha (q=1)=0$.}, and $\alpha
(q=0.8)=0.15$ and $\alpha (q=0.6)=0.34$, in agreement with the values given
in Tab. \ref{Tab1}. In addition, fitting the numerical values of $\gamma $
also with a second-order polynomial function of $q$, we have found that $%
\gamma (q)=0.25\,q^{2}-0.25\,q+1.75$, for $0.5<q\leq 1.0$ (see Fig.~\ref%
{Fig5}), or 
\begin{equation}
\gamma (q)=\frac{1}{4}(q^{2}-q+7),  \label{eq.12b}
\end{equation}%
which give us the exact known value $\gamma (q=1)=\frac{7}{4}$,  $\gamma
(q=0.8)=1.69$ and $\gamma (q=0.6)=1.71$. These results are also close to the
ones obtained numerically. On the other hand, fitting the numerical values
of $\beta $ with a straight line, we have found that $\beta
(q)=-0.124+0.249\,q$, for $0.5<q\leq 1.0$ (see Fig.~\ref{Fig5}). We may
conjecture that the exact dependence of $\beta $ on $q$ in this range is 
\begin{equation}
\beta (q)=\frac{1}{8}(2q-1),  \label{eq.12c}
\end{equation}%
which give us the exact known value $\beta (q=1)=\frac{1}{8}$, $\beta
(q=0.8)=0.075$ and $\beta (q=0.6)=0.025$, values that are also close to ones
the obtained numerically.

These results suggest a nonuniversality of the critical exponents along the
ferromagnetic-paramagnetic frontier. In addition, it also suggest that the scaling relations 
\begin{eqnarray}  \label{eq.13}
2\,\beta+\gamma & = & d\,\nu ~, \nonumber \\
\alpha+2\,\beta+\gamma & = & 2 ~, 
\end{eqnarray}
\noindent where $d$ is the dimension of the lattice ($d=2$ for the square
lattice), should be changed. Thus, if we consider the above dependence of $%
\alpha$, $\beta$ and $\gamma$ on $q$, the first scaling relation of Eqs.~(\ref%
{eq.13}), will become 
\begin{eqnarray}  \label{eq.14}
2\,\beta+\gamma & = & (d+n_{q})\,\nu ~,
\end{eqnarray}
\noindent where 
\begin{eqnarray}  \label{eq.15}
n_{q} & = & \frac{1}{4}\left(q^{2} + q - 2\right) ~.
\end{eqnarray}
\noindent Notice that for $q=1$, one has $n_{1} = 0$, and the
standard scaling relation is recovered. On the other hand, although $\alpha$, $\beta$, 
and $\gamma$ depend on the entropic index $q$, the Rushbrooke equality is satisfied for all
$0.5 < q \leq 1.0$, within uncertainty.


\section{Conclusions}

We have studied the Ising model with nearest-neighbors interactions on a
square lattice by means of numerical Monte Carlo simulations. In our
approach, different from other authors \cite%
{1999_PhysA_268_553,2000_PhysA_283_59,1996_PhysA_233_395,1997_PhysA_242_250,1997_PhysA_247_553}%
, we simply changed the weight in the Metropolis algorithm to a ratio
between the escort probabilities of the nonextensive statistics. This study
was motivated by possible connection of the Tsallis statistics and some
manganese oxides, called manganites, like $\mathrm{%
La_{0.60}Y_{0.07}Ca_{0.33}MnO_{3}}$ \cite%
{2006_PRB_73_092401,2002_PRB_66_134417,2002_EPL_58_42,2003_PRB_68_014404}.
Due to computational cost, our simulations were done after $10^{7}$ Monte
Carlo steps, with the entropic index $q \in [0,1]$ and the linear lattice
sizes $L = 32, 64, 128, 256$ and $512$.

The Monte Carlo simulation of an Ising model with nearest-neighbors
interactions showed a distinct behavior of the same system considered in the
infinite-range-interaction limit \cite{2003_PRB_68_014404}. Jumps on the
magnetization and susceptibility curves in the range $0.0<q<0.5$ occur in
both approaches, but for short-range interactions we do not have first-order
phase transitions. In addition, the mean-field calculations foresee the same
critical exponents of the 2D Ising model in the framework of the
Boltzmann-Gibbs statistics. However, our calculation of the magnetization,
the susceptibility and the specific heat for the short-range interacting
system showed that three of the critical exponents depend on $q$ in the
range $0.5<q\leq 1.0$.

Finite-size scaling analysis of the results showed that the critical
exponents $\alpha$, $\beta$ and $\gamma$, that are related to the behavior
of the specific heat, the magnetization and the susceptibility near the
critical point $T_{c}$, respectively, depend on $q$ in the range $0.5< q\leq
1.0$. Based on the numerical estimates of these exponents, we conclude that the dependencies are of the form $%
\alpha(q)=(10\,q^{2}-33\,q+23)/20$, $\beta(q)=(2\,q-1)/8$ and $%
\gamma(q)=(q^{2} - q + 7)/4$. Although the exponents $\alpha$, $\beta$ and $%
\gamma$ depend on $q$, as well as the critical temperatures $T_{c}$ \cite%
{2008_EPJB_62_337}, the exponent $\nu$ does not; we found that $\nu=1.0$ $%
\forall$ $q$. These dependencies of the critical exponents on the entropic
index suggest a nonuniversality of those exponents along the
ferromagnetic-paramagnetic frontier. It also suggest a violation of 
the scaling relations $\alpha+2\,\beta+\gamma = 2$ (Rushbrooke equality) and 
$2\,\beta+\gamma=d\,\nu$. However, when we take into account the $q-$dependence 
of the critical exponents showed in Table~\ref{Tab1}, we notice that the former 
scaling relation should be changed to $2\,\beta + \gamma = (d + n_{q})\,\nu $, 
where $n_{q} = \left(q^{2} + q - 2\right)/4$ (note that for $q = 1$, we obtain 
$n_{1} = 0$), but the Rushbrooke equality is not altered. Thus, the inhomogeneities 
introduced in the system by the nonextensive statistics may be responsible for the $q-$dependence of the
critical exponents $\alpha$, $\beta$ and $\gamma$, as well as the critical
temperatures $T_{c}$.



On the other hand, we have a completely different scenario in the
range $0.0 <q\leq 0.5$. The cutoff of the Tsallis distribution keep the
system in the ground state (with $m=1$) for $T < T_{c}=4(1-q)$, and at $T_{c}$ the
magnetization jumps suddenly to zero, i.e., to a equiprobable state
\cite{2008_EPJB_62_337}. In the same way, the susceptibility and the specific heat curves
also present jumps at $T_{c}$, due to the cutoff. Although the presence of
these jumps, the histograms of the energy states visited during the dynamics, 
at the critical temperatures, show only one-peak structures, which is a 
indicative of the occurrence of continuous phase transitions.

Previous works on long- and short-range interactions 1D Ising models \cite%
{1999_PRL_83_4233,2000_EPJB_17_679,2001_PhysA_290_159} predict that the
magnetization scales differently for $q<1.0$ and $q=1.0$ regimes. Therefore,
in this work we showed that the magnetization of the short-range 2D Ising
model scales also differently in two regimes: for $0.5 < q \leq 1.0$ the
system scales as a 2D Ising model, but for $q \leq 0.5$ the magnetization
and the critical temperature are independent of the lattice size due to the
cutoff; thus, the scaling appears naturally on the system.

Also in a previous work \cite{1999_PhysA_268_553}, it was shown that 2D
Ising model with nearest-neighbors interactions does not undergo phase
transitions, except for $q=1.0$. The main difference between their approach
and ours is related to the definition of the temperature scale. In that work 
\cite{1999_PhysA_268_553}, the authors have chosen $\beta $ as the parameter
related to the temperature scale and, in this work, we have chosen $\beta
_{q}^{\prime }$. The relation between these parameters is given in Eq.(\ref%
{eq.04}). The advantage of our approach over previous one, for the choice of
the temperature scale, is that ours is supported by previous description of
the magnetic properties, experimentally and theoretically investigated, of
manganites \cite%
{2006_PRB_73_092401,2006_EPJB_50_99,2002_PRB_66_134417,2002_EPL_58_42,2003_PRB_68_014404}%
. Thus, based on that, we believe that the 2D Ising model undergoes a phase
transition even for $q\neq 1.0$, and the scaling relations should be changed
as described above.

Extensions of this work to describe inhomogeneous magnetic systems, i.e.,
systems in which the exchange interaction changes along the sites of the
lattice, as well as the study of the effects of uniform and random magnetic
fields, within a nonextensive approach would be of great interest, because
it can yield some clues to questions about the connection of such systems
and the nonextensive statistics.

\vskip 2\baselineskip

{\large \textbf{Acknowledgments}}

\vskip \baselineskip
\noindent

The authors acknowledge S.M.D. Queirós for his comments. We would like to
thanks the Brazilian funding agencies CNPq, CAPES and the Brazilian
Millennium Institute for Quantum Information for the financial supports.
D.O.S.P. thanks FAPESP for financial support, M.S.R. thanks the financial
support from PCI-CBPF program and A.M.S. would like to thanks the Ontario
Goverment.

\vskip 2\baselineskip

\end{document}